\documentclass[aps,reprint,twocolumn,superscriptaddress,citeautoscript,showpacs,amsmath,amssymb,floatfix,prb,longbibliography]{revtex4-1}
\usepackage{}
\usepackage{graphicx}
\usepackage{amsmath}
\usepackage{amssymb}
\usepackage{mathtools}
\usepackage{textcomp,gensymb} 
\usepackage{bm} 
\usepackage[usenames,dvipsnames]{xcolor}
\usepackage{float}
\usepackage{setspace}
\usepackage{dcolumn}
\usepackage{array}

\usepackage{changepage}
\usepackage{tikz}
\usepackage{contour}
\usepackage[left]{lineno}
\usepackage{blindtext}
\usepackage{setspace}
\usepackage[none]{hyphenat}
\usepackage[colorlinks=true,citecolor=blue,linkcolor=blue,pdftex,hypertexnames=false,urlcolor=blue,linktocpage=true]{hyperref}
\definecolor{ashgray}{rgb}{0.7,0.75,0.71}
\definecolor{mspringgreen}{rgb}{0, 0.8, 0.1}
\definecolor{auburn}{rgb}{0.43, 0.21, 0.1}
\definecolor{ao(english)}{rgb}{0.0, 0.5, 0.0}
\definecolor{afw}{rgb}{0.95, 0.95, 0.96}
\definecolor{magnolia}{rgb}{0.97, 0.96, 1.0}
\definecolor{wsmk}{rgb}{0.96, 0.96, 0.96}

\newcolumntype{M}[1]{>{\centering\arraybackslash}m{#1}}

\usepackage{dcolumn}
\usepackage{multirow} 
\usepackage{soul}
\usepackage{titlesec}
\usepackage{color}

\usepackage{bbold}

\newcolumntype{d}[1]{D{.}{.}{#1}}

\begin{document}

\title{Tuning the structure and superconductivity of SrNi$_2$P$_2$ by Rh substitution.}

\author{J. Schmidt}
\affiliation{Department of Physics and Astronomy, Iowa State University, Ames, IA 50011, USA}
\affiliation{Ames National Laboratory, Iowa State University, Ames, IA 50011, USA}
\author{A. Sapkota}
\affiliation{Ames National Laboratory, Iowa State University, Ames, IA 50011, USA}
\author{C. L. Mueller}
\affiliation{Department of Physics and Astronomy, Iowa State University, Ames, IA 50011, USA}
\author{S. Xiao}
\affiliation{Department of Materials Science and Engineering and Institute of Materials Science, University of Connecticut, 97 North Eagleville Road, Unit 3136, Storrs, Connecticut 06269-3136, USA}
\author{S. Huyan}
\affiliation{Department of Physics and Astronomy, Iowa State University, Ames, IA 50011, USA}
\affiliation{Ames National Laboratory, Iowa State University, Ames, IA 50011, USA}
\author{T. J. Slade}
\affiliation{Ames National Laboratory, Iowa State University, Ames, IA 50011, USA}
\author{R. A. Ribeiro}
\affiliation{Department of Physics and Astronomy, Iowa State University, Ames, IA 50011, USA}
\affiliation{Ames National Laboratory, Iowa State University, Ames, IA 50011, USA}
\author{S.-W. Lee}
\affiliation{Department of Materials Science and Engineering and Institute of Materials Science, University of Connecticut, 97 North Eagleville Road, Unit 3136, Storrs, Connecticut 06269-3136, USA}
\author{S. L. Bud'ko}
\affiliation{Department of Physics and Astronomy, Iowa State University, Ames, IA 50011, USA}
\affiliation{Ames National Laboratory, Iowa State University, Ames, IA 50011, USA}
\author{P. C. Canfield}
\affiliation{Department of Physics and Astronomy, Iowa State University, Ames, IA 50011, USA}
\affiliation{Ames National Laboratory, Iowa State University, Ames, IA 50011, USA}

\date{\today }

\pacs{1234}

\begin{abstract}

SrNi$_2$P$_2$ is unique among the ThCr$_2$Si$_2$ class since it exhibits a temperature induced transition upon cooling from an uncollapsed tetragonal (ucT) state to a one-third-collapsed orthorhombic (tcO) state where one out of every three P-rows bond across the Sr layers. This compound is also known for exhibiting bulk superconductivity below 1.4 K at ambient pressure. In this work, we report on the effects of Rh substitution in Sr(Ni$_{1-x}$Rh$_x$)$_2$P$_2$ on the structural and superconducting properties. We studied the variation of the nearest P-P distances as a function of the Rh fraction at room temperature, as well as its temperature dependence for selected compositions. We find that increasing the Rh fraction leads to a decrease in the transition temperature between the ucT and tcO states, until a full suppression of the tcO state for $x\geq 0.166$. The superconducting transition first remains nearly insensitive to the Rh fraction, and then it increases to 2.3 K after the tcO state is fully suppressed. These results are summarized in a phase diagram, built upon the characterization by energy dispersive x-ray spectroscopy, x-ray diffraction, resistance, magnetization and specific heat measurements done on crystalline samples with varying Rh content. The relationship between band structure, crystal structure and superconductivity is discussed based on previously reported band structure calculations on SrRh$_2$P$_2$. Moreover, the effect of Rh fraction on the stress-induced structural transitions is also addressed by means of strain-stress studies done by uniaxial compression of single-crystalline micropillars of Sr(Ni$_{1-x}$Rh$_x$)$_2$P$_2$.

\end{abstract}

\maketitle

\section{Introduction} 
\label{sec:Introduction}

The family of compounds with $AM_2X_2$, with $A$ an alkali, alkali earth or rare earth metal, $M$ a transition metal, and $X$ a $p$-block element, is a well known composition frequently adopting the ThCr$_2$Si$_2$ structure. The compounds in this family have attracted abundant interest, offering a diversity of tunable structural, magnetic, electronic and superconducting properties \cite{Szytula1994,Gati2020,Canfield2009,Ni2008a,Gati2012,Trovarelli2000,Budko1999,Avila2004,Kong2014}. Among this family, some members display structures where the distance between nearest $X$ atoms is compatible with the presence of $X$-$X$ bonds across the layers of $A$ atoms \cite{Hoffmann1985}. The term collapsed tetragonal (cT) was coined \cite{Kreyssig2008} to describe the greatly reduced $X$-$X$ distance associated with this bonding state, whereas those that do not present $X$-$X$ bonding are referred to as uncollapsed tetragonal (ucT). This terminology was extended to closely related structures that do not strictly belong ThCr$_2$Si$_2$ category: the CaKFe$_4$As$_4$ family of materials \cite{Iyo2016}, most commonly known as 1144, can adopt a ucT state at low pressures, a cT state at high pressures, and a half-collapsed tetragonal (hcT) state at intermediate pressures in which As atoms bond across every other $A$-layer \cite{Kaluarachchi2017,Borisov2018,Xiang2018b,Xiang2022,Stillwell2019,Wang2023}; and the one-third-collapsed orthorhombic (tcO) state \cite{Schmidt2023} can be adopted by SrNi$_2$P$_2$ in which only one of every three P-P rows bonds across the Sr layers when cooled down below $325\ \text{K}$ \cite{Barth1997}.

The $X$-$X$ distance in these systems is highly tunable by different means such as applying pressure \cite{Kreyssig2008,Gati2012,Kaluarachchi2017,Borisov2018,Torikachvili2008,Yu2009}, uniaxial strain \cite{Xiao2021,Sypek2017}, chemical substitution \cite{Ran2014,Jia2009} and thermal treatments \cite{Ran2011,Ran2014}. This has allowed for experimental access to a collapsed tetragonal transition in which the $X$-$X$ bonds form/break, and for studies of how this impacts the mechanical \cite{Sypek2017,Bakst2018,Xiao2021}, superconducting \cite{Kaluarachchi2017,Gati2012} and magnetic \cite{Kreyssig2008,Canfield2009,Jia2011} properties of these materials. In addition, these transitions give rise to a remarkable pseudoelasticity \cite{Sypek2017,Bakst2018}, which allows some of these materials to achieve some of the highest maximum recoverable strains, yield strength, and modulus of resilience among metals \cite{Xiao2021}, which make them potentially interesting for engineering applications \cite{Juan2009,Huber1997}. 

The interplay between structural degrees of freedom and superconductivity has been widely explored in compounds with  ThCr$_2$Si$_2$-related structures, as well as in compounds with completely different structures. For example, the tetragonal (but nearly cubic) BaBi$_3$ can undergo first-order phase transitions into different structures upon applying pressure, leading to clearly distinct superconducting states for each structure \cite{Xiang2018}; and PbTaSe$_2$ can also undergo a pressure-induced first-order structural phase transition that involves the displacement of the Pb atoms leading to a step-like decrease of the superconducting transition temperature \cite{Kaluarachchi2017b}. Studies of Ca(Fe$_{1-x}$Co$_x$)$_2$As$_2$ show that the application of hydrostatic pressure can induce a transition from the ucT state to a cT state, with a simultaneous change from superconducting to non superconducting states \cite{Gati2012}. Similarly, the pressure induced transition from the ucT to hcT state in CaKFe$_4$As$_4$ and its derivatives results in the full suppression of superconductivity \cite{Kaluarachchi2017,Xiang2018,Xiang2022}. More recently, the effect of the collapsed tetragonal transition on superconductivity was studied for the BaTi$_2$(Sb$_{1-x}$Bi$_x$)$_2$O system $\cite{Yajima2013}$. Upon the application of hydrostatic pressure, a collapsed tetragonal transition is induced, resulting in the formation of pnictogen-pnictogen bonds $\cite{Yamamoto2021,Ikeda2022}$. It has been observed that the superconducting transition temperature is enhanced when inducing the collapsed tetragonal phase by applying pressure $\cite{Ikeda2022}$. This behavior is opposite to the rest of the cases mentioned in which the superconductivity thrives only in the ucT phase.

SrNi$_2$P$_2$ offers a unique opportunity to explore the effect of collapse tetragonal transitions involving tcO states on superconductivity, as it has been reported to be superconducting for temperatures below 1.4 K \cite{Ronning2009}. Although the relationship between the tcO$\leftrightarrow$ucT transition and magnetism has recently been addressed \cite{Schmidt2023}, the relationship between the tcO phase and superconductivity is still widely unexplored. Since the tcO structure is intermediate between the ucT and cT structures, SrNi$_2$P$_2$ allows us to study the changes in the superconducting properties due to a transition between ucT and tcO states and between the tcO and cT states. Previous work done on Sr$_{1-x}$Ba$_x$Ni$_2$P$_2$ has demonstrated that increasing the Ba content can enhance the superconducting transition temperature to 2.85 K in the vicinity of the transition between tcO and ucT states, which is higher than those of the end members SrNi$_2$P$_2$ and BaNi$_2$P$_2$ (1.4 K and 2.5 K, respectively).$\cite{Kudo2017}$ However, due to limitations of that work, discussed in the Section \ref{sec:structure_superconductivity}, the behavior of the superconducting transition temperature for those compositions corresponding to the tcO phase is unclear. 

Motivated by this, we explore the case of Sr(Ni$_{1-x}$Rh$_x$)$_2$P$_2$. As Rh is added ($x$ increases from zero) the ucT to tcO structural phase transition temperature ($T_S$) decreases until reaching 0 K for $0.122<x<0.166$. Despite previous reports stating that there is a coexistence of tcO and ucT phases well below the transition temperature \cite{Xiao2021,Schmidt2023}, we show that only the tcO phase is present in single crystals with $x<0.122$ at low temperatures. We report on the effects of Rh substitution on the structural properties, including the tuning of $T_S$, and its the effect on the stress induced collapse transitions. Moreover, we focus on the changes in the superconducting properties that occur after Rh substitution fully suppresses the tcO state and fully stabilizes the ucT state.

\section{Experimental Details}
\label{sec:Experimental}

Single crystals of Sr(Ni$_{1-x}$Rh$_x$)$_2$P$_2$ were obtained by the high-temperature solution growth method \cite{Canfield2020}
out of Sn flux \cite{Schmidt2023} for $x\leq0.098$, and out of self flux for $x\geq 0.122$. The pure elements were loaded into a $2\ \text{ml}$ alumina fritted Canfield Crucible Set \cite{CanfieldP.C.KongT.KaluarachchiU.S.2016,LSPCeramics}, and sealed under partial atmosphere of argon in a fused silica tube. For the crystals grown out of Sn flux, starting compositions of Sr$_{1.3}$(Ni$_{1-x}$Rh$_x$)$_2$P$_{2.3}$Sn$_{16}$ were used. The ampoules were placed inside a box
furnace, held for 4 hours at $600\ ^{\circ}\text{C}$ before increasing to $1150\ ^{\circ}\text{C}$, held for 24 hours to make sure the material was fully melted, and finally slowly cooled down over 250 h to $650\ ^{\circ}\text{C}$, at which the excess Sn was decanted with the aid of a centrifuge \cite{Canfield2020}. 

When attempting the Sn-flux growth with an initial composition of Sr$_{1.3}$(Ni$_{1-x}$Rh$_x$)$_2$P$_{2.3}$Sn$_{16}$ with $x\geq0.7$ (nominal), crystals of Sr$_3$Sn$_2$P$_4$ phase were grown instead of the targeted Sr(Ni$_{1-x}$Rh$_x$)$_2$P$_2$. The Sn-flux method was only effective on crystals with nominally $x\leq 0.6$, which correspond to an actual Rh content of $x\leq 0.098$ on the crystals, according to Energy Dispersive x-ray Spectroscopy (EDS) measurements (see below). The self-flux method was used for growing samples with higher Rh fraction.

For the crystals grown out of self-flux, Sr$_5$(Ni$_{1-x}$Rh$_x$)$_{48}$P$_{47}$ was used as the ratio for the pure elements placed in the alumina crucible \footnote{A previous step was performed in order to \textit{own the eutectic} of (Ni$_{1-x}$Rh$_x$)$_{60}$P$_{40}$ as explained in reference [\!\!\!\citenum{Slade2022}].}. Powder Ni and Rh were used, and the temperature was increased to $1190\ ^{\circ}\text{C}$ with a slow rate in order to avoid large vapour pressures of P, as explained in reference [\!\!\!\citenum{Slade2022}]. A two-step fractionation process was used \cite{Slade2022}, similar to the one reported for growing CePd$_3$S$_4$ \cite{Huyan2024} or EuPd$_3$S$_4$ \cite{Huyan2023}: (1) the melt was slowly cooled down to $1050\ ^{\circ}\text{C}$, at which the mixture was decanted in order to separate out the grown RhP$_2$ crystals; (2) the decanted side was resealed into a new silica ampoule, heated to $1190\ ^{\circ}\text{C}$, fast cooled down to $1100\ ^{\circ}\text{C}$ and finally slowly cooled down to $1000\ ^{\circ}\text{C}$. At this temperature the mixture was decanted, and plate-like crystals of Sr(Ni$_{1-x}$Rh$_x$)$_2$P$_2$ were obtained. Depending on the composition, the decanted temperature had to be adjusted to slightly higher temperatures (up to $1025\ ^{\circ}\text{C}$) in order to stay above the eutectic temperature.

When attempting the self-flux growth method for $x<0.15$ (nominal), the crystals obtained gave diffraction patterns consistent with SrNi$_5$P$_3$ instead of the target phase. Most likely, the Ni in this crystals is partially substituted with Rh, which does not significantly affect the diffraction pattern, and would require EDS measurements for confirmation. Additionally, when attempting this method for $x>0.35$ (nominal) tetragonal crystals of Ni-substituted Sr$_2$Rh$_7$P$_6$ were obtained. The self-flux method only succeeded for nominal values of $x$ intermediate to these, which gave crystals with actual Rh fractions in the range $0.122\leq x\leq0.245$, according to EDS measurements.

Single crystals of SrRh$_2$P$_2$ ($x=1$) were obtained by extended grain growth out of a Pb flux. Pure metals were loaded into a $2\ \text{ml}$ alumina fritted Canfield Crucible Set \cite{CanfieldP.C.KongT.KaluarachchiU.S.2016,LSPCeramics} with the composition SrRh$_2$P$_2$Pb$_{20}$, and sealed under partial atmosphere of Argon in a fused silica tube. The ampoules were placed inside a box
furnace, held for 4 hours at $600\ ^{\circ}\text{C}$ before increasing to $1200\ ^{\circ}\text{C}$, at which they were held for 24 hours. At this temperature the excess Pb was decanted with the aid of a centrifuge. Plate-like single crystals of SrRh$_2$P$_2$ with submilimetric dimensions.

The Rh concentration levels ($x$) were determined by Energy Dispersive x-ray Spectroscopy (EDS) quantitative chemical analysis with an EDS detector (Thermo NORAN Microanalysis System, model C10001) attached to a JEOL scanning-electron microscope (SEM). An acceleration voltage of
$22\ \text{kV}$, working distance of $10\ \text{mm}$ and take off angle of $35 ^{\circ}$ were used for measuring all standards and crystals with unknown composition. Single crystals of SrNi$_2$P$_2$ and of SrRh$_2$P$_2$ were used as a standards for Sr, Ni, Rh and P quantification. The spectra were fitted using NIST-DTSA II Microscopium software \cite{Newbury2014}. The composition of each platelike crystal was measured on at least four different positions on the crystal's face (perpendicular to $c$). The crystals with $x_{\text{nominal}}\leq 0.4$ were thick enough in order to polish their edges and measure the composition on at least three different points across the edge of the crystal (along $c$) as well, whereas those for higher Rh fraction were too thin to allow this type of polishing without breaking. The average compositions and error bars were obtained from these data, accounting for both inhomogeneity and goodness of fit of each spectra. It should be noted that, unlike the Sr(Ni$_{1-x}$Co$_x$)$_2$P$_2$ system \cite{Schmidt2023}, the Sr(Ni$_{1-x}$Rh$_x$)$_2$P$_2$ system does not suffer from strong inhomogeneity along the $c$-axis. 

Powder x-ray diffraction (XRD) measurements were performed using a Rigaku MiniFlex II powder diffractometer with Cu-K$\alpha$ radiation ($\lambda=1.5406\ \text{\AA}$). For each composition, a few crystals were finely ground to powder and dispersed evenly on a single crystal Si zero background holder, with the aid of a small quantity of vacuum grease. Intensities were collected for $2\theta$ ranging from $15^{\circ}$ to $100^{\circ}$, in step sizes of $0.01^{\circ}$, counting for 4 seconds at each angle. Rietveld refinement was performed on each spectra using GSAS II software package \cite{Dreele2014}. Refined parameters included but were not limited to phase fractions, lattice parameters, atomic positions and isotropic displacements. The same equipment was used to measure single-crystal samples which were cleaved along the
(001) plane, following the method described in reference \citenum{Jesche2016}. 

Single crystal x-ray diffraction was performed using a
Rigaku XtaLab Synergy-S diffractometer with Ag radiation ($\lambda=0.56087\ \text{\AA}$), in transmission mode, operating at 65 kV and 0.67 mA. The samples were held in a nylon loop with Apiezon N grease. The temperature was controlled using Oxford Cryostream 100, by adjusting the flow of cold nitrogen gas on the crystal. The transition temperatures extracted from these single crystal x-ray diffraction measurements are consistent, within 3 K, with those extracted from other measurement techniques. The total number of runs and images was based on the strategy calculation from the program CrysAlisPro (Rigaku OD, 2023). The data integration and reduction were also performed using CrysAlisPro, and a numerical absorption correction was applied based on Gaussian integration over a face-indexed crystal. The structures were solved by intrinsic phasing using the SHELXT software package and were refined with SHELXL.

 The temperature dependent AC resistance of the samples was measured using a Quantum Design Physical Property Measurement System (PPMS) using the AC transport (ACT) option, with a frequency of $17\ \text{Hz}$ and a $3\ \text{mA}$ excitation current. Only in-plane resistance was measured, using a standard four-contact geometry. Electrical contacts with less than $1.5\ \Omega$ resistance were achieved by spot welding $25\ \mu\text{m}$ Pt wire to the samples, followed by adding Epotek H20E silver epoxy, and curing the latter for 1 hour at $120^{\circ}\text{C}$. In selected cases, the base temperature of the PPMS was extended down to 130 mK using a Cambridge ADR module for PPMS from CamCool
Research Limited. Nominal zero field measurements were collected as the temperature drifted back to the PPMS base temperature.

DC magnetization measurements were carried out on a Quantum Design Magnetic Property Measurement System (MPMS classic) superconducting quantum interference device (SQUID) magnetometer (operated in the range $1.8\ \text{K}\leq T \leq 350\ \text{K}$, at a field of $H = 10\ \text{kOe}$). Both zero field cooling (ZFC) and field cooling (FC) protocols were used. Each sample was measured with the field applied parallel and perpendicular to the tetragonal $c$ axis. The samples were glued on a Kel-F disk which was placed inside a plastic straw; the contribution of the disk to the measured magnetic moment was independently measured in order to subtract it from our results. 

Temperature-dependent specific heat measurements were carried out in a Quantum Design DynaCool PPMS using the relaxation technique as implemented in the heat capacity option, using heat pulses corresponding to a 2\% temperature raise, and fitting the temperature relaxation with a two-$\tau$ model. Measurements were performed under no applied magnetic field.

For the study of the mechanical properties of the crystals, micropillars were produced using an FEI Helios Nanolab 460F1 focused ion beam (FIB) machine as described in reference \citenum{Sypek2017}. The in situ nanomechanical tests were performed at room temperature and under ultra-high vacuum condition using NanoFlip (Nanomechanics, Inc., TN, USA), which is installed in a field-emission gun JEOL 6330F scanning electron microscope.

\section{Results and discussion}
\label{sec:Results}
\subsection{Elemental analysis}

\begin{figure}[tbh]
 \includegraphics[width=\linewidth]{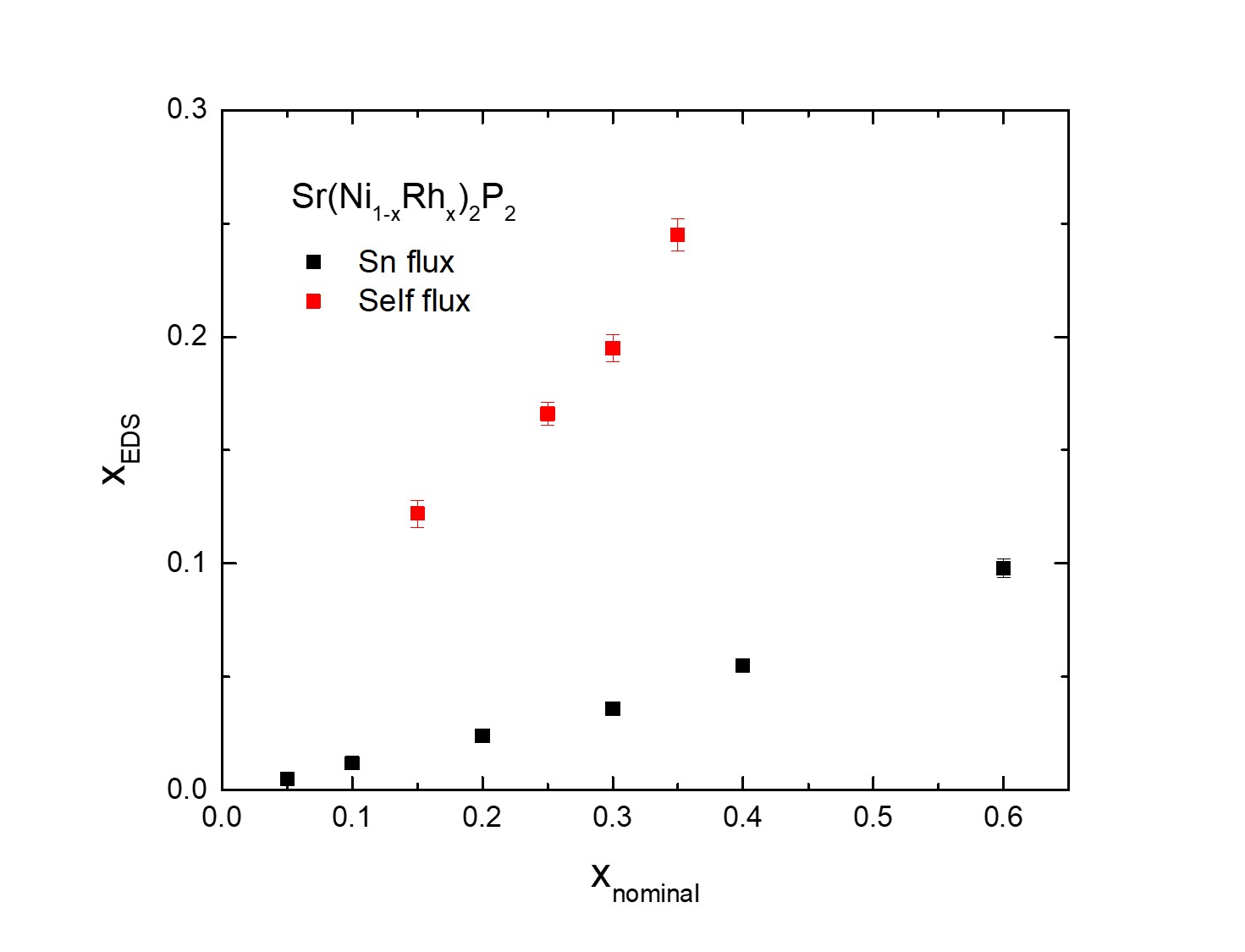}
 \caption{\footnotesize{Fraction of Rh in Sr(Ni$_{1-x}$Rh$_x$)$_2$P$_2$} determined by EDS as a function of the nominal Rh fraction that was originally present in the high temperature melt that crystals were grown out of.}
 \label{fig:EDS}
\end{figure}

The Rh substitution determined by EDS, $x_{\text{EDS}}$, is shown in Fig. \ref{fig:EDS} as a function of the nominal Rh fraction, $x_{\text{nominal}}$, that was originally used to create the high temperature solution out of which the crystals were grown. The black symbols correspond to those compositions grown out of a Sn flux, whereas the red symbols correspond to those that were grown out of a self-flux. It can be easily noted that the self-flux technique allows for significantly larger Rh fractions than the Sn-flux. The uncertainty due to the goodness of fit of the EDS spectra was the main contribution to the error bars presented in the figure, and the inhomogeneneities in the distribution of Rh, if present, were considerably smaller. From this point in the text, the composition referred to corresponds to that measured by EDS, and the symbol $x$ would be used instead of the full $x_{EDS}$ in order to simplify notation.

\begin{figure}[tbh]
 \includegraphics[width=\linewidth]{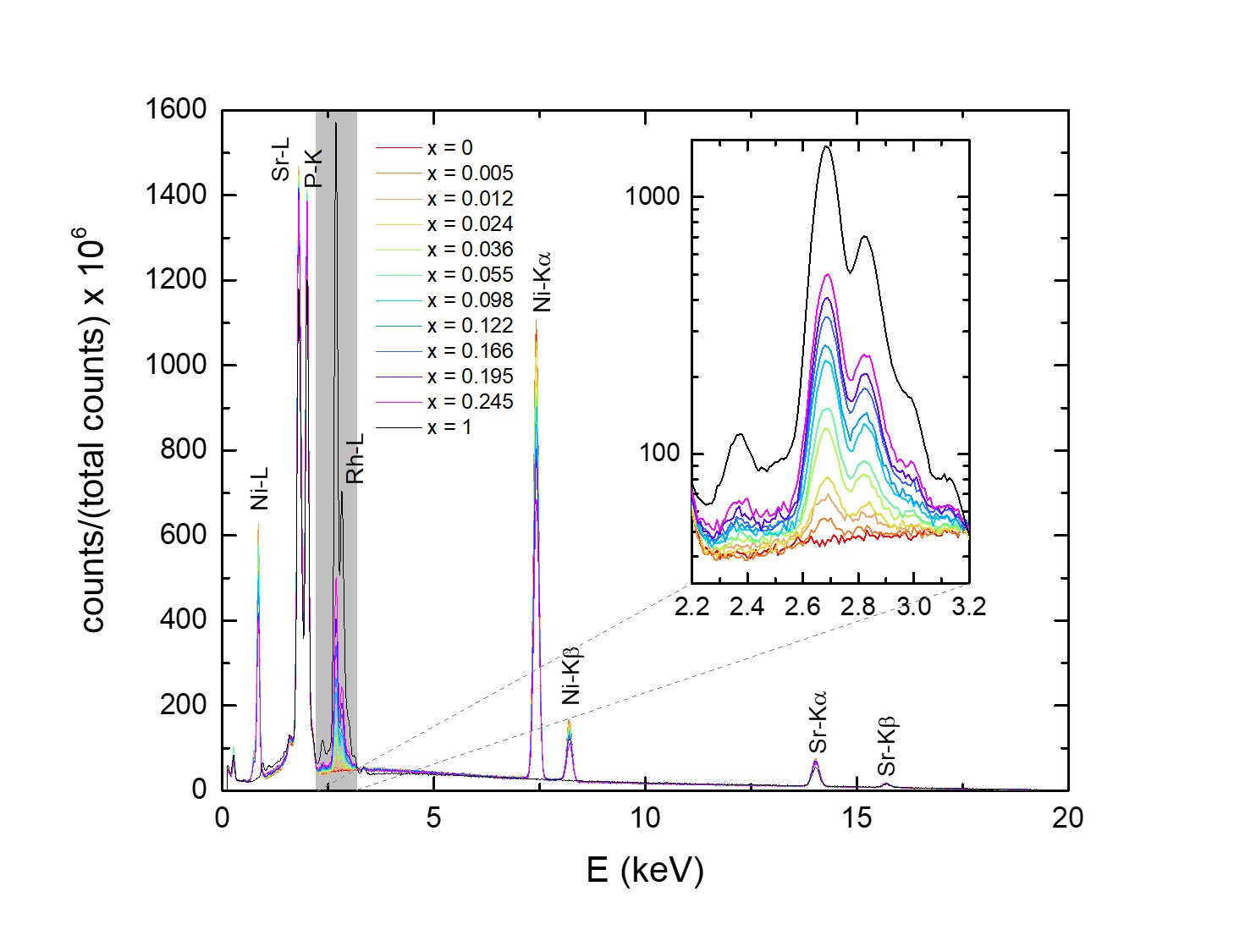}
 \caption{\footnotesize{Main text: EDS spectra for Sr(Ni$_{1-x}$Rh$_x$)$_2$P$_2$} with different Rh fractions. Inset: semi-logarithmic plot of the Rh-$L$ peaks.}
 \label{fig:EDS_spectra}
\end{figure}

The EDS spectra for the samples of Sr(Ni$_{1-x}$Rh$_x$)$_2$P$_2$ with different values of $x$ are plotted in Fig. \ref{fig:EDS_spectra}. The inset shows the enhancement of the Rh-$L$ peaks, qualitatively consistent with the increasing Rh fraction $x$. A semi-logarithmic scale was used to visualize simultaneously the Rh-$L$ peaks for all values of $x$ explored in this work. Even the composition with least Rh fraction ($x=0.005$) shows a discernible peak that differs significantly from the pure SrNi$_2$P$_2$.

\subsection{Effects of Rh substitution on the crystal structure}

Single crystal x-ray diffraction was performed on crystals of each composition at room temperature. Tables \ref{tab:scxrd_SrNi2P2}-\ref{tab:scxrd_SrRh2P2} in Appendix A show the information of the single crystal refinements on Sr(Ni$_{1-x}$Rh$_x$)$_2$P$_2$ for $x=0$, $x=0.024$, $x=0.098$ and $x=1$, for room temperature. Twins, typical for orthorhombic subgroups of tetragonal groups, were observed and refined for $x=0$, where the $a$ and $b$ axes of one variant are exchanged with respect to the other variant. This occurs due to the ucT$\leftrightarrow$tcO transition above room temperature, in which tripling of the unit cell occurs along any of the two equivalent axis of the tetragonal high temperature structure.

\begin{figure}[tbh]
\centering
\includegraphics[width=\linewidth]{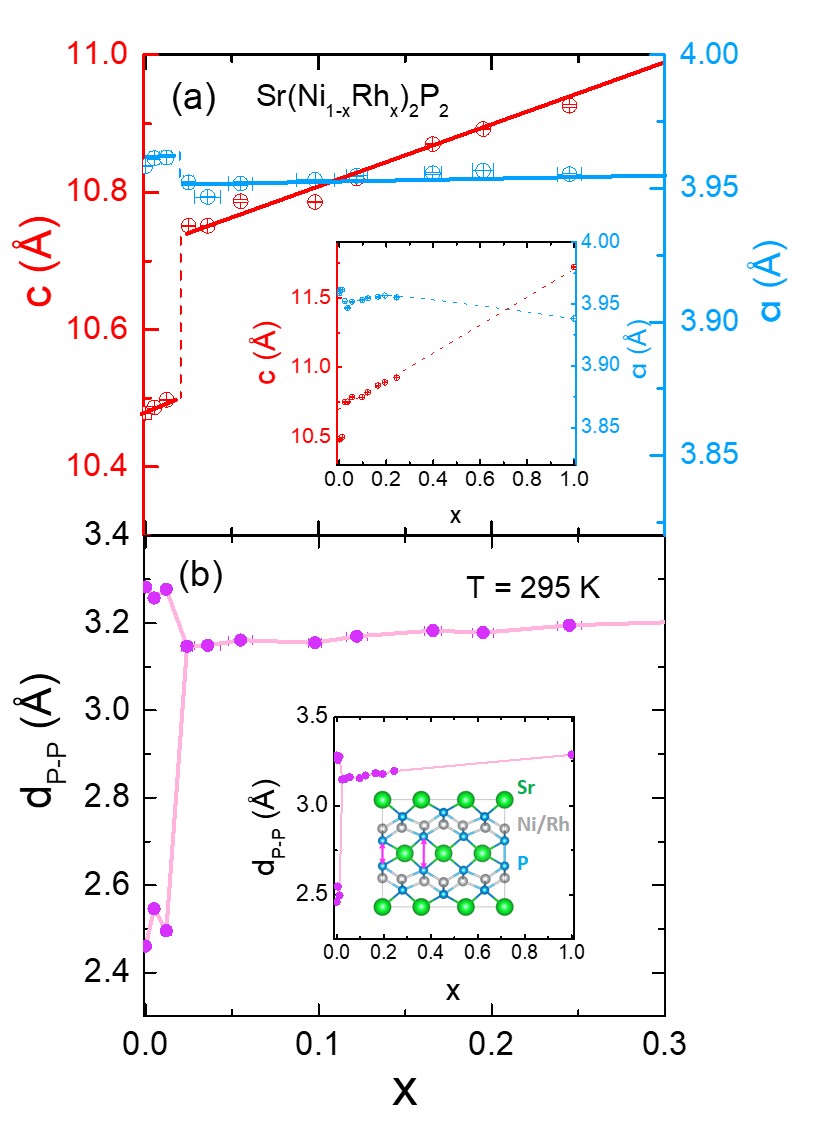}
\caption{\footnotesize{(a) Dependence of the room temperature $c$-lattice parameter (red, left axis) and $a$-lattice parameter (blue, right axis) of Sr(Ni$_{1-x}$Rh$_x$)$_2$P$_2$ as a function of $x$, measured on single crystals. Solid and dashed lines were added as guides to the eye. The inset shows the whole composition range $0\leq x\leq 1$, and the main panel shows a smaller $0\leq x\leq 0.3$. (b) P-P distances across the Sr plane as a function of $x$, as indicated with purple arrows on the drawing of the structure shown in the inset. The green atoms correspond to Sr, blue to P and grey to Ni or Rh.}}
\label{fig:SC_latticeparam}
\end{figure}

By solving and refining each structure, the room temperature lattice parameters, atomic positions and space group were found. The two determined crystal space groups are orthorhombic $Immm$ for the tcO phase and $I4/mmm$ for the ucT phase. The main panel in Fig. \ref{fig:SC_latticeparam}(a) shows the dependence of the room temperature $c$ (red, left axis) and $a$ (blue, right axis) lattice parameters with the Rh fraction $x$. The values of $b$ are not displayed given that they are equal to $a$ for the ucT phase, and to 3$a$ for the tcO phase. A clear upward jump of approximately 2\% in $c$ is observed between $x=0.012$ and $x=0.024$ as the room temperature phase changes from tcO to ucT, whereas the change in $a$ is only of 0.2\%. Fig. \ref{fig:SC_latticeparam}(b) shows the P-P distances across the Sr plane (as indicated by the purple arrows in the drawing of the crystal structure in the inset) at room temperature as a function of the Rh fraction $x$. In the tcO phase (at small Rh concentrations), there are two distinct P-P distances corresponding to the bonded (shorter) and unbonded (longer) P-P pairs. The shortest P-P distance across the Sr plane increases by approximately 20\% whereas the other, longer inter-plane P-P distance shows a smaller decrease, with the two becoming equal as the ucT phase is stabilized at room temperature for $x\geq 0.024$. As shown in both insets, the trends in the $c$-lattice parameter and the P-P distances of the substitution series reasonably extrapolate to those corresponding to SrRh$_2$P$_2$, whereas the $a$-axis data imply that there is a weak non-monotonic behavior.

All of these results obtained by single crystal x-ray diffraction were consistent with a single phase in all the crystals measured. This contrasts with previous reports of coexistence of ucT and tcO phases in pure SrNi$_2$P$_2$ \cite{Xiao2021,Schmidt2023}, even far beyond the region of hysteresis reflected in temperature dependent resistance measurements. In Appendix B, we compare the results obtained through different diffraction techniques in order to argue that the coexistence observed previously was most likely induced by the stress exerted to the crystals when grinding them or applying other perturbations on them.

\begin{figure}[H]
\centering
\includegraphics[width=\linewidth]{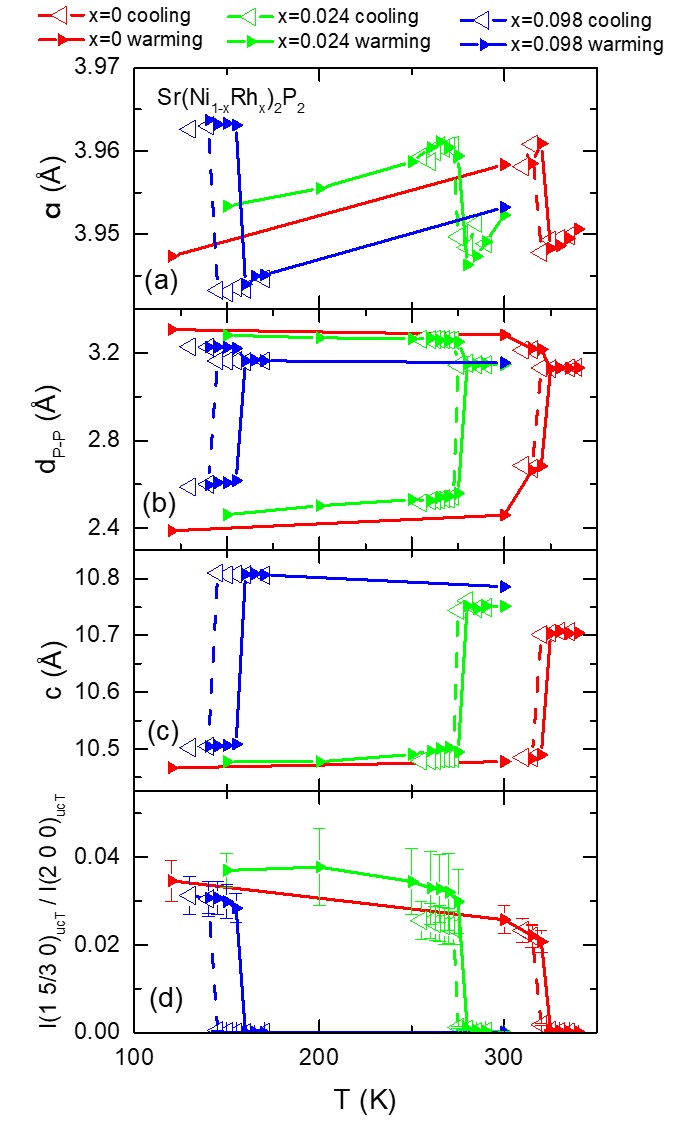}
\caption{\footnotesize{(a) $a$-lattice parameter, (b) P-P distances across the Sr plane, (c) $c$-lattice parameter, and (d) the ratio between the (1 5/3 0) satellite peaks (indexed in the ucT basis) and (200) nuclear peaks, as a function of temperature. Open triangle symbols pointing left connected by dashed lines correspond to the data measured upon cooling, and solid triangle symbols pointing right connected by solid lines correspond to the data measured upon warming.}}
\label{fig:scxrd}
\end{figure}

Additionally, single crystal x-ray diffraction measurements were performed at different temperatures for $x=0$, $x=0.024$ and $x=0.098$, in order to study the effects of Rh substitution on the ucT to tcO transition. Fig. \ref{fig:scxrd}(b) shows there is a sudden decrease in nearest P-P distance, $d_{P\text{-}P}$, upon cooling below a structural transition temperature, $T_S$, due to the formation of P-P bonds across the Sr layers as the system transitions from the ucT phase to the tcO phase. A smaller increase in the remaining P-P distances occurs at the same temperature. Figs. \ref{fig:scxrd}(a) and (c) present the evolution of the lattice parameters as a function of temperature, showing that upon cooling below $T_S$, the $a$-lattice parameter has a sudden increase and the $c$-lattice parameter has a sudden decrease, while the opposite occurs upon warming. Whereas the relative change in one of the $d_{P-P}$ separation is approximately 20\%, it is important to recall that this represents only one third of the P-P separations across the Sr layer. As such, there is only a 2\% decrease in the $c$-lattice parameter.

A hysteresis of a few K between cooling and warming can be appreciated in all the transitions, consistent with a first-order nature of the transition. The transition temperatures decrease upon increasing the Rh fraction. The size of the jump in both $a$ and $c$ increases with Rh doping (and decreasing $T_S$ value), whereas the size of the jump in the P-P distances decreases. This again highlights the complexity of having multiple P-P distances across the Sr plane in the tcO phase.

Fig. \ref{fig:scxrd}(d) plots the ratio between one type of tcO satellite peaks, appearing at $ (1\ 5/3\ 0)$ with respect to the tetragonal I4/mmm, and a nuclear peak common to ucT and tcO structures as a function of temperature. The sudden jump in intensity of the satellite peaks from zero to a finite value upon cooling is consistent with the first order nature of the transition, as already mentioned. The fact that it is zero at temperatures above the transition (beyond the range of hysteresis) indicates that there is no metastable tcO phase coexisting with the ucT phase at higher temperatures. This supports the idea that the coexistence observed by the powder x-ray diffraction measurements for $x=0.024$ shown in Fig. \ref{fig:plate_powder}(d) is only a consequence of the stress exerted to the crystals when grinding. 

\begin{figure}[tbh]
\centering
\includegraphics[width=\linewidth]{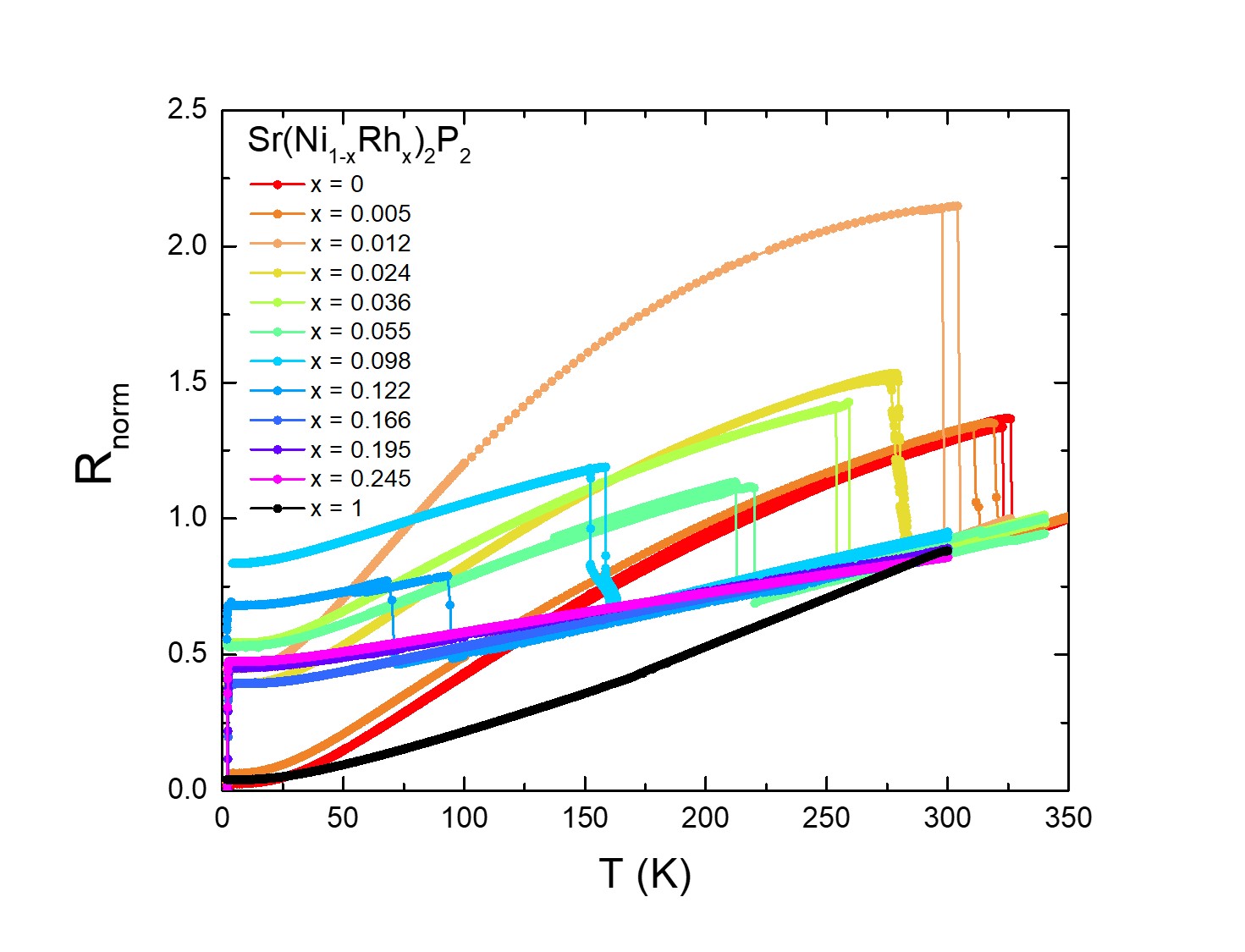}
\caption{\footnotesize{Temperature dependent resistance upon cooling and warming, normalized to its value at $350\ \text{K}$, $R(T)/R(350\ \text{K})$, for samples with different $x$. $R(T)$ curves for $x=0.098$, $x=0.122$, $x=0.166$, $x=0.195$, $x=0.245$ and $x=1$ are normalized to the value of $R($300 K$)$ of $x=0.055$.}}
\label{fig:rt}
\end{figure}

\subsection{Effects of Rh substitution on the electronic and magnetic properties}

 Fig. \ref{fig:rt} presents the normalized resistance as a function of temperature for samples with different $x$. A sharp step can be identified both upon lowering and increasing the temperature, which corresponds to the sudden tcO $\leftrightarrow$ ucT structural transition. The temperatures of both steps (upon cooling and upon warming) differ, giving rise to a hysteretic behavior, consistent with a first order phase transition. The transition temperatures and widths of hysteresis were consistent with the temperature dependent x-ray diffraction results shown in Fig. \ref{fig:scxrd}.  Fig. \ref{fig:rt} shows that $T_S$ decreases with increasing $x$, dropping below 2 K between $x=0.122$ and $x=0.166$. It should be noted that, for $x=0.012$, room temperature is within the width of hysteresis, so it would not be a surprise to find some coexistance of ucT and tcO phases at room temperature, even without grinding. This however was not observed by single crystal x-ray diffraction measurements.

\begin{figure}[tbh]
\centering
\includegraphics[width=\linewidth]{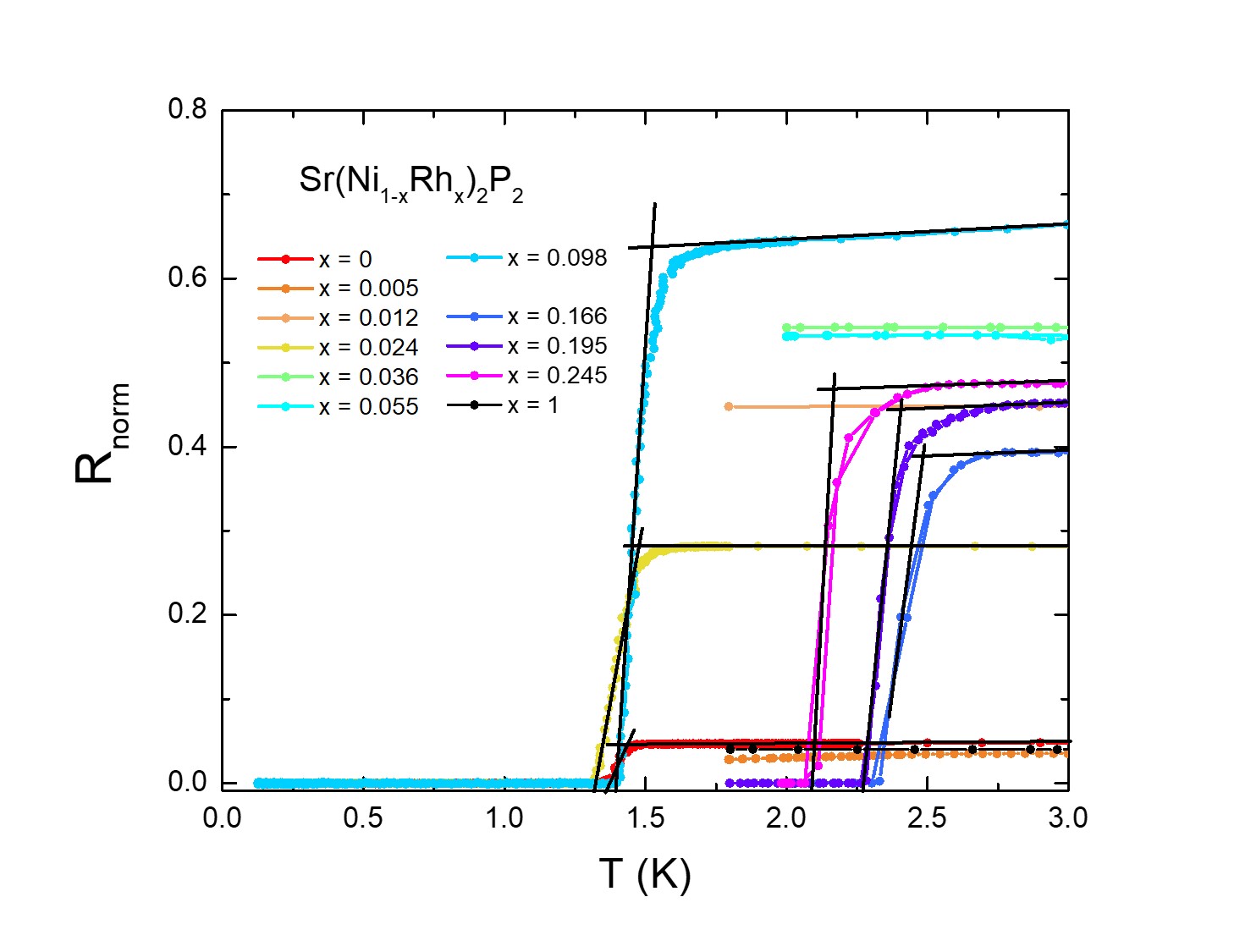}
\caption{\footnotesize{Temperature dependent resistance upon cooling and warming, for the range 0 K $<$ $T$ $<$ 3 K, normalized to its value at $350\ \text{K}$, $R(T)/R(350\ \text{K})$, as done in Fig. \ref{fig:rt}. Only four representative samples were measured to temperatures below 1.8 K. It should be noted that the lack of data  for lower temperatures does not indicate a lack of superconductivity for other compositions with $x < 0.122$. The black lines indicate the lines of maximum slope and the extrapolated high temperature behavior, used to estimate the onset and offset temperatures of the transition.}}
\label{fig:rt_sc}
\end{figure}

Fig. \ref{fig:rt_sc} shows an enlarged view of the lowest temperatures at which resistance was measured, in order to be able to observe the drop to zero resistance due to the superconducting transition. Most of the curves were measured down to a minimum temperature of 1.8 K. Only a four \textit{representative} samples ($x=0$, 0.024, 0.098 and 0.122) were measured to lower temperatures by means of adiabatic demagnetization refrigeration in order to confirm the presence of superconductivity below 1.8 K. The $T_c$ in this case was determined as the offset of the transition, by intersecting the line of maximum slope (shown in black) with the temperature axis. The onset of the transition was also estimated from the intersection between the line of maximum slope and the extrapolated high temperature behavior. It can be seen that the superconducting transition temperature $T_c$ is nearly independent of the Rh fraction for $x\leq0.098$ and it is enhanced to temperatures above 2 K for $0.166\leq x\leq 0.245$. 

The low temperature resistance for $x=0.122$ measured on four different samples is plotted in Fig. \ref{fig:RT_ZX312} in Appendix C. It displays a more variable $T_c$ as well as a broadened superconducting transition, despite the similarity in the values of $x$ of each particular sample. The estimated onsets range from 1.9 K to 2.9 K, and the differences between each onset and the corresponding offset are much larger than for the other compositions, reaching to 1 K for the sample A plotted in Fig. \ref{fig:RT_ZX312}. This is consistent with $x=0.122$ corresponding to the lowest $T_S$ measured, very close to the point where the tcO transition is suppressed to 0 K (see Fig. \ref{fig:phase_diagram} below). It is also worth noting that $x=0.122$ has a wider temperature range in which either the ucT or tcO phases remain metastable, as evidenced with the larger width of hysteresis of the structural transition temperature in Fig. \ref{fig:rt}. It is possible that some metastable ucT phase persists in some parts of the sample even at the lowest measured temperatures, which would be expected to become superconducting at higher temperatures than the parts that transform into the tcO state. This could explain the broader and more complicated shape of the superconducting transition, and the greater dispersion of the onset and offset of superconductivity compared to the rest of the compositions. As shown in Fig. \ref{fig:Mag_sc_zx312} in Appendix C, the magnetic susceptibility at 1.8 K is consistent with less than 1\% of the sample being superconducting at that temperature. However, even a small superconducting fraction of the sample can lead to a notable decrease in the resistance, especially if it can percolate throughout the sample.

\begin{figure}[tbh]
\centering
\includegraphics[width=\linewidth]{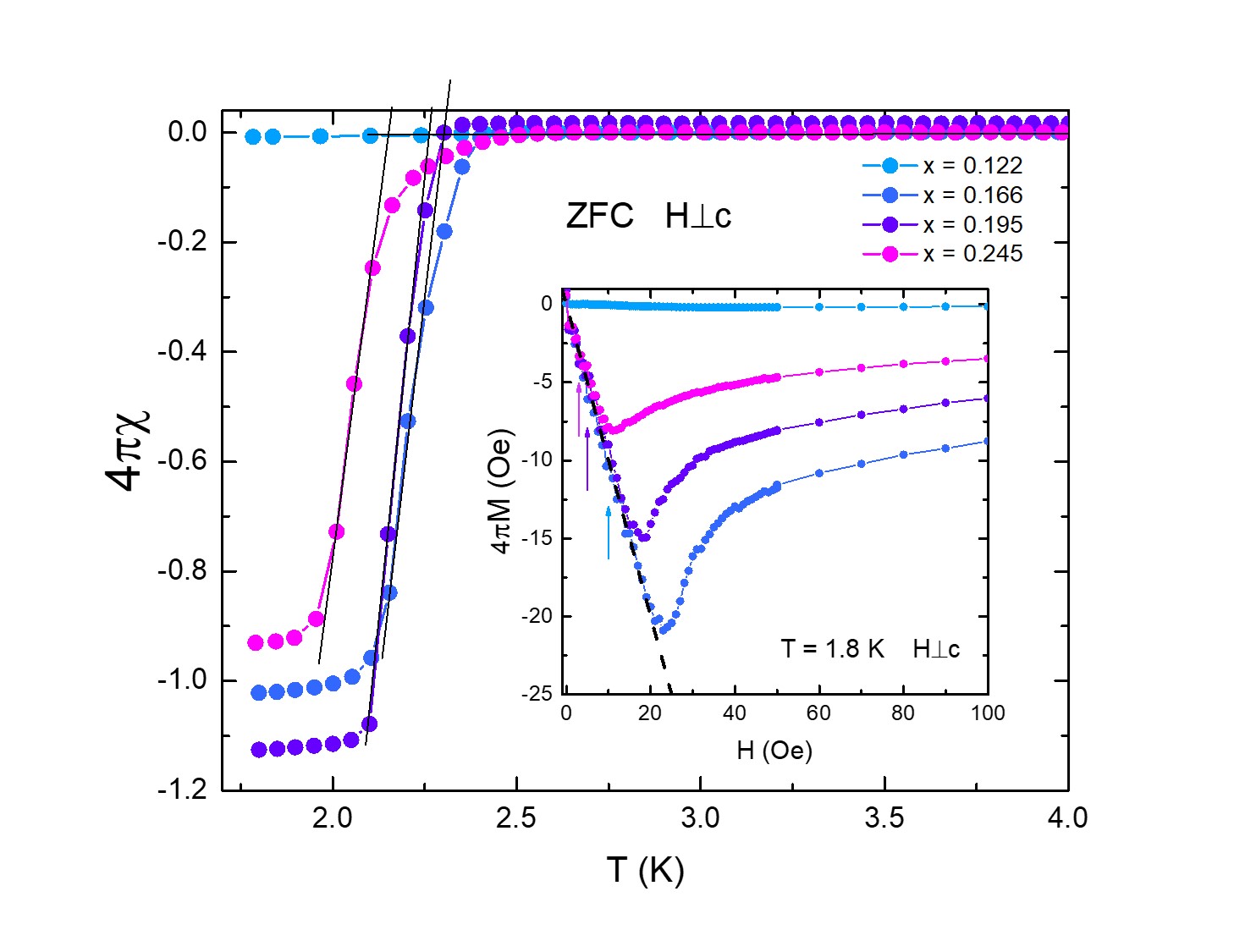}
\caption{\footnotesize{Main panel: magnetic susceptibility as a function of temperature for Sr(Ni$_{1-x}$Rh$_x$)$_2$P$_2$ with $0.122\leq x\leq 0.488$. Inset: Magnetization as a function of magnetic field for the former compositions. The susceptibility curves plotted in the main panel were measured at fields within the linear regime shown in the inset, and are indicated by the arrows of colors of the corresponding curves. }}
\label{fig:Mag_sc}
\end{figure}

Zero-field cooled DC magnetic susceptibility $\chi$ measurements were done to further characterize the superconductivity in the samples with $x=0.166$, $x=0.195$ and $x=0.245$, as shown in the main panel of Fig. \ref{fig:Mag_sc}. The saturation of $4\pi\chi$ to values close to $-1$ at low temperatures indicates a nearly complete expulsion of the magnetic field from the entire volume of the sample. The samples used for these measurements have a plate-like shape, where the dimensions along the $a$ and $b$-axes are at least ten times larger than the dimension along the $c$-axis. These were measured with an applied field perpendicular to the $c$-axis, which results in demagnetization factors that are lower than 0.1. Hence, the product of the demagnetization factor and the magnetization values are much smaller than the applied magnetic fields, and can be ignored. The intersection of the black lines sketched in the main panel was used as the criteria for determining the $T_c$, which gave consistent values to those obtained from the resistance measurements. The magnetic field dependence of the zero-field cooled magnetization was also measured for $T=1.8$ K, reflecting the linear behavior of $4\pi M$ with a slope of $-1$ for $H\lesssim H_{c1}$, and then an inverted peak shape followed by an increase in the magnetization for higher fields, as expected for type II superconductors. These results suggest that upon increasing the Rh fraction above $x=0.166$ there is a gradual suppression of $T_c$, which eventually drops below 1.8 K for SrRh$_2$P$_2$. This qualitatively resembles to the case of Sr$_{1-x}$Ba$_x$Ni$_2$P$_2$ where there is a less pronounced suppression of the $T_c$ from 2.85 K to 2.5 K as the Ba fraction is increased for the ucT phase.$\cite{Kudo2017}$

The superconductivity in SrNi$_2$P$_2$ has been argued to be of BCS nature, through fitting of the heat capacity measurements \cite{Ronning2009}, and exponential behavior of the thermal conductivity \cite{Kurita2009}. The persistence (and even enhancement) of superconductivity in Sr(Ni$_{1-x}$Rh$_x$)$_2$P$_2$ for $x\geq 0.166$ after the ucT state is fully stabilized is consistent with predictions that Ni substitution could induce superconductivity in SrRh$_2$P$_2$ \cite{Johrendt1997}. This claim was based on the similarity of the band structure of SrRh$_2$P$_2$ with that of LuNi$_2$B$_2$C, both with similar crystal structures and with a peak in  density of states due to a Van Hove singularity in the $M-M$ antibonding bands ($M$=Rh or Ni). This peak in the density of states lies at the Fermi energy for LuNi$_2$B$_2$C and has been associated with the superconductivity in this compound, and it lies about 0.2 eV above the Fermi energy for the non-superconducting SrRh$_2$P$_2$. For this reason, substitution with Ni, Pd or La has been suggested as ways to increase the Fermi energy in order to match the peak in the density of states and induce superconductivity. It should be noted that this prediction does not clarify how much Ni is needed to tune the Fermi energy towards this peak in the density of states, and that the compositions explored in this work contain Ni fractions that range from 0.755 to 1, when viewing them as Ni substituted SrRh$_2$P$_2$. Furthermore, the predictions in reference \citenum{Johrendt1997} are limited to compositions in the ucT state, not addressing if this conjecture for superconductivity is compatible with compositions that have a tcO groundstate.

\begin{figure}
\centering
\includegraphics[width=\linewidth]{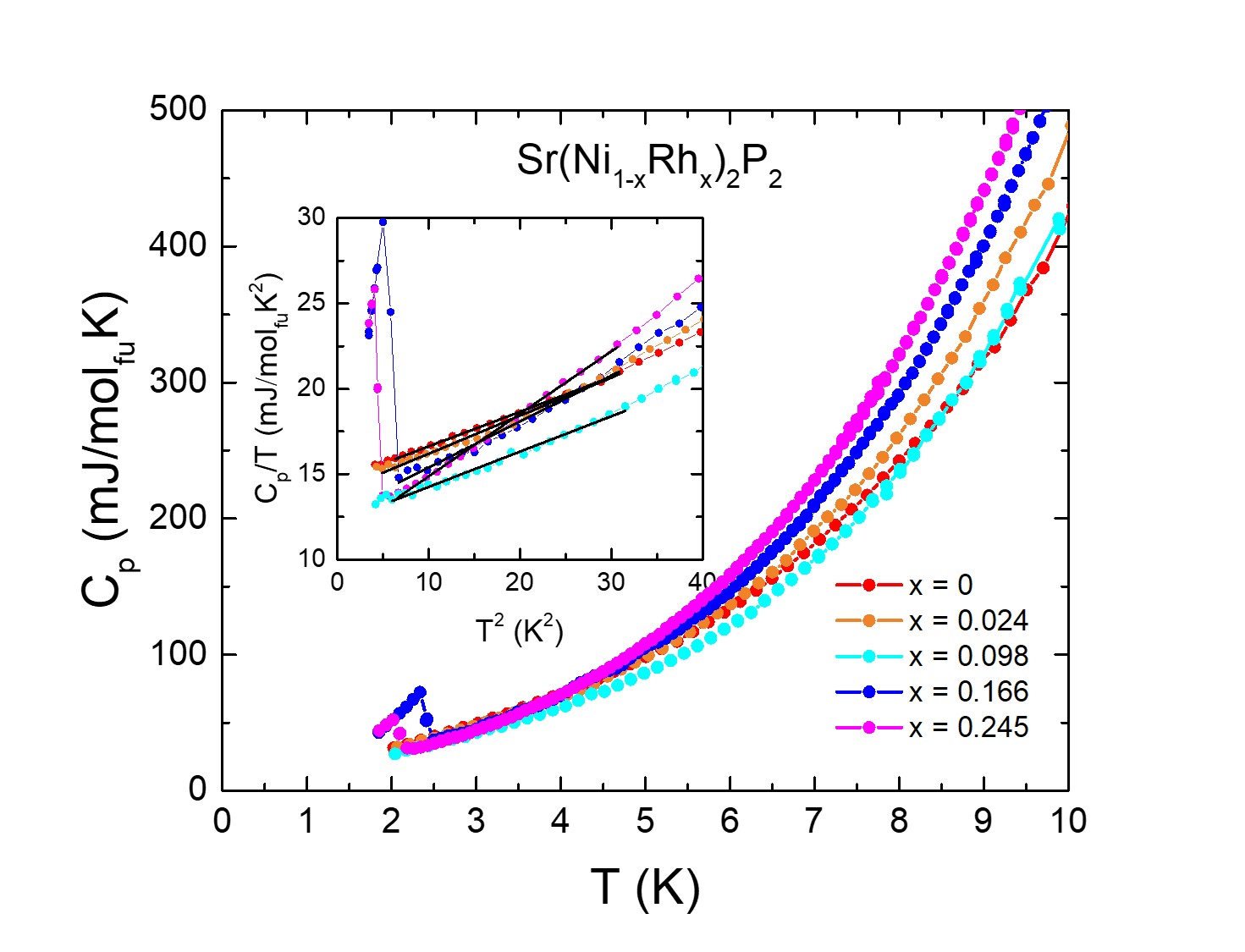}
\caption{\footnotesize{Main panel: Temperature dependence of the specific heat, $C_p$, in units of mJ/K per mole of formula unit, for $x=0$, 0.024, 0.098, 0.166 and 0.245. Inset: $C_p/T$ of Sr(Ni$_{1-x}$Rh$_x$)$_2$P$_2$ as a function of the temperature squared. The linear fits are represented with black lines.}}
\label{fig:cp}
\end{figure}

The main panel of Fig. \ref{fig:cp} shows the low temperature behavior of the specific heat, $C_p$, for selected compositions. Since the measurements were done for $1.8\ \text{K}\leq T\leq 10\ \text{K}$, the jump in specific heat associated with the superconducting transition of the samples with $x<0.122$ was not observed, as they have $T_c<$ 1.8 K. The jumps associated to the superconducting transition for $x=0.166$ and $x=0.245$ were sharp enough to easily distinguish its onsets and offsets by eye. Their $T_c$ was defined as the average between the onset and the offset, and the uncertainty as half of the difference between them. These were consistent with those determined by resistance and magnetization measurements, as can be seen in Fig. \ref{fig:phase_diagram} and Table \ref{tab:phase_diagram}. 

The Sommerfeld coefficient, $\gamma$, which is proportional to the density of states at the Fermi level, can be inferred from the inset of Fig. \ref{fig:cp} as it is defined as the intercept of the linear fit of $C_p/T$ vs $T^2$. The values obtained for $\gamma$ are in the range $11-15$ mJ/mol$_{fu}$K$^2$, which is comparable to that of LuNi$_2$B$_2$C, which was reported \cite{Michor1995} to be of 19.5(3) mJ/mol$_{fu}$K$^2$. The size of the specific heat jump, $\Delta C_p$ was also quantified for $x=0.166$ and $x=0.245$ obtaining values of $38(2)\ $mJ/mol$_{fu}$K and $27(2)\ $mJ/mol$_{fu}$K, respectively. This yielded values of $\Delta C_p/\gamma T_c$ equal to 1.23(7) and 1.15(7), for $x=0.166$ and $x=0.245$, respectively. These values are slightly lower than the expected 1.43 expected for BCS theory, as it occurs in numerous other conventional superconductors \cite{Ronning2009,Bauer2008,Klimczuk2009,Kasahara2008}.

\begin{figure}
\centering
\includegraphics[width=\linewidth]{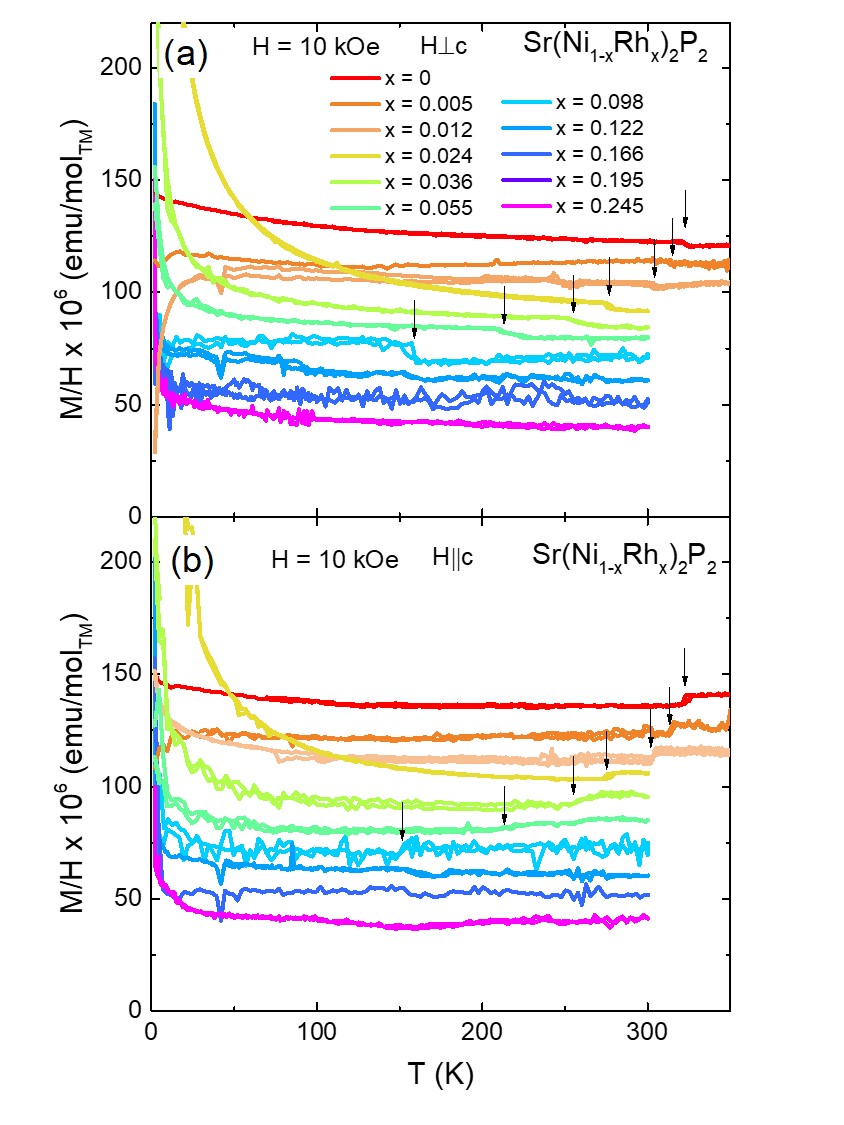}
\caption{\footnotesize{Temperature dependent magnetization normalized by the applied magnetic field, measured upon warming and cooling, in units of emu per transition metal atom, for the applied field perpendicular to $c$ (a) as well as parallel to $c$ (b). Manual offsets were added for clarity of the data.}}
\label{fig:mt}
\end{figure}

Higher field and higher temperature anisotropic magnetization data, in units of emu per mole of transition metal atom and normalized by the applied magnetic field, are presented in Fig. \ref{fig:mt} for samples with different compositions. The data was collected by applying a magnetic field of $H=$ 10 kOe, perpendicular (upper panel) as well as parallel (lower panel) to the crystallographic $c$-axis. In order to allow for a clearer comparison of the features displayed for each composition, an offset was added to the different curves. The results for $H\perp c$ show a step-like feature consistent with the transition temperature determined by resistance measurements (see Fig. \ref{fig:rt}). A more subtle step can be appreciated in some of the magnetization measurements for $H||c$ as well, although $M_{tcO,||}<M_{ucT,||}$ as opposed to the data for $H \perp c$ which clearly shows that $M_{tcO,\perp}>M_{ucT,\perp}$. These results are consistent with previously reported results on SrNi$_2$P$_2$ \cite{Ronning2009}, and indicate a change in the anisotropy of the electronic properties. Furthermore, these results show no sign of magnetic orders emerging upon suppressing the structural collapse transition temperature, in contrast to what was observed in Sr(Ni$_{1-x}$Co$_x$)$_2$P$_2$ as well as other related systems \cite{Schmidt2023,Jia2009,Jia2011}. This does not rule out the possibility of magnetic ordering occurring in samples with higher Rh fractions than the ones that we were able to grow for this work.

\subsection{Interplay between structure and superconductivity}
\label{sec:structure_superconductivity}

\begin{figure}[H]
 \centering
 \includegraphics[width=\linewidth]{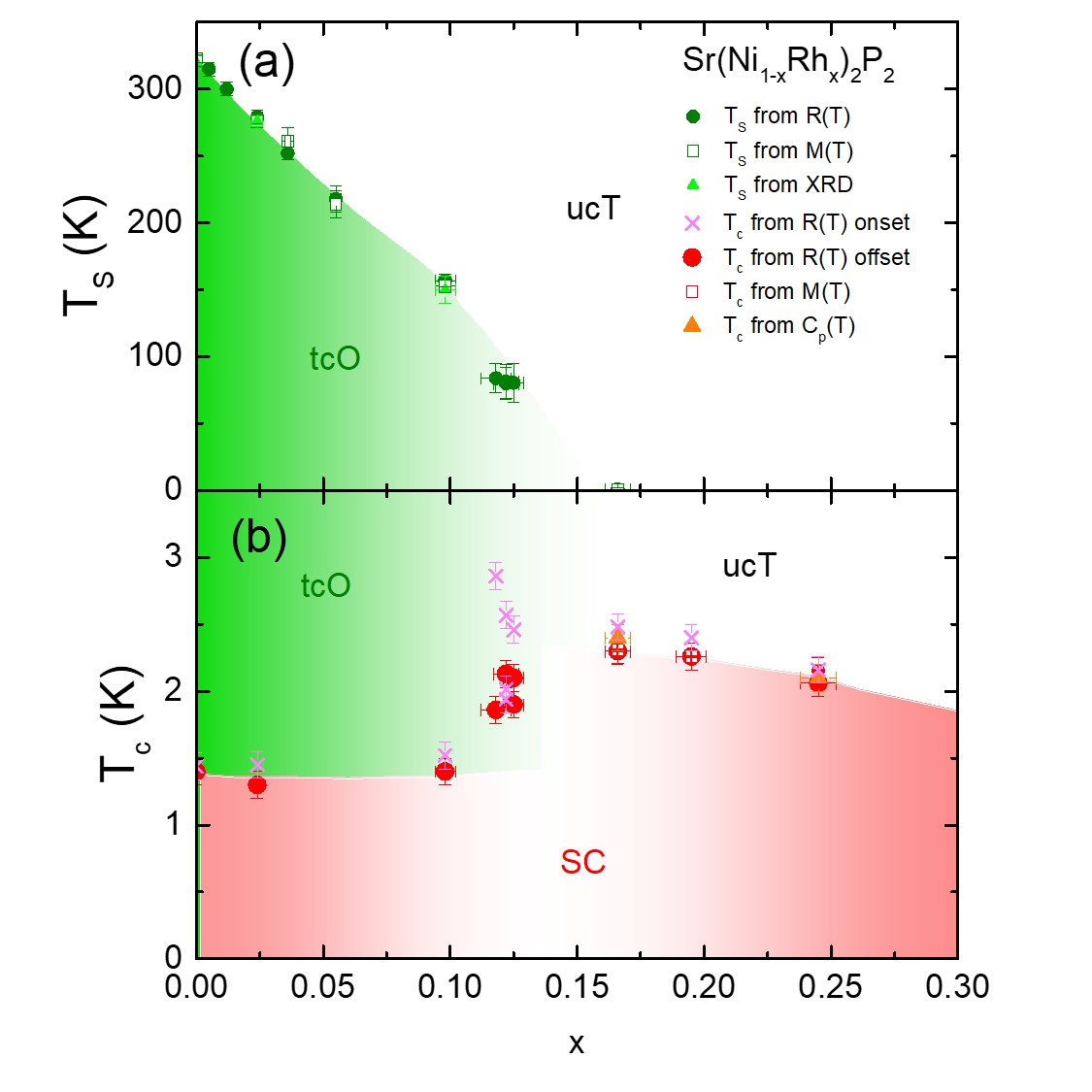}
 \caption{\footnotesize{(a) Structural transition temperature, $T_S$, as a function of the Rh fraction, $x$. The solid green symbols correspond to the temperatures obtained from the resistance data, the open symbols correspond to those obtained from magnetization measurements with field applied perpendicular to $c$, and the light-green triangles correspond to the temperatures obtained from single crystal x-ray diffraction. The green region corresponds to the tcO phase. (b) Superconducting transition temperature, $T_c$, as a function of the Rh fraction, $x$. The pink crosses correspond to the onset temperatures obtained from resistance data, the solid red circles correspond to the offset temperatures obtained from resistance data, the open red squares correspond to the temperatures obtained from magnetization data, and the orange triangles correspond to the temperatures obtained from specific heat data. The red region corresponds to the superconducting phase. The lack of red symbols for $x<0.122$ (with the exception of $x=0$, 0.024 and 0.098) does not indicate a lack of superconductivity, but rather that measurements were not performed below 1.8 K where superconductivity is expected.}}
 \label{fig:phase_diagram}
\end{figure}

The ucT $\leftrightarrow$ tcO structural transition temperature, $T_S$, as well as the superconducting transition temperature, $T_c$, inferred from the above data are plotted as a function of the Rh fraction, $x$, in Fig. \ref{fig:phase_diagram} as a $T-x$ phase diagram. Since there is hysteresis associated to the structural phase transition, the onset temperatures of the transition upon cooling and upon warming were averaged in order to obtain an estimate of the structural transition temperature, $T_S$, and their difference was taken as an estimation of the uncertainty of $T_S$. These results are also summarized in Table \ref{tab:phase_diagram}. This explicitly shows the suppression of the structural transition with Rh substitution, until only the ucT phase is present for the whole temperature range in the samples with $x\geq0.166$. The $T_c$ on the other hand, remains nearly independent of $x$ for those compositions with a tcO groundstate, and it is enhanced in the ucT phase after the full suppression of the tcO state. The results for the samples with $x=0.122(6)$ display a wide variety of onset and offset temperatures for superconductivity, which may be an indication of possible metastable coexistance of tcO and ucT phases, with lower and higher $T_c$ values, respectively.

Both Rh and Ba substitution in SrNi$_2$P$_2$ result in a comparable enhancement of the $T_c$ near the boundary that separates the ucT and the tcO phases, reaching 2.4 K for Sr(Ni$_{1-x}$Rh$_x$)$_2$P$_2$ and 2.85 K for Sr$_{1-x}$Ba$_x$Ni$_2$P$_2$ $\cite{Kudo2017}$. However, the behavior of $T_c$ for the values of $x$ corresponding to the tcO phase of the latter system is unclear: even though the onset of the transition remains at 2.85 K, the superconducting volume fraction is less than 50\%. It should be noted that the mentioned work does not report any EDS results or any alternative technique to measure the real compositions of the samples, but rely only on the nominal composition used for the synthesis. Therefore, it is possible that the Ba could be inhomogeneously distributed, and that the apparently enhanced superconductivity in the tcO state is due to portions of the sample with high enough Ba content ($x\geq 0.5$) to be in the ucT state. Alternatively, even with a homogeneous distribution of Ba, it could be that the superconductivity observed for $0.12\leq x\leq 0.4$ is due to some metastable ucT phase, as was observed in this work for Sr(Ni$_{1-x}$Rh$_x$)$_2$P$_2$ with $x=0.122(6)$. 

The relationship between the crystal structure and band structure in materials with collapsed tetragonal structure has been addressed and discussed in previous works \cite{Hoffmann1985,Borisov2018,Johrendt1997}. The formation of P-P bonds was originally attributed to the depopulating of a P-P antibonding band \cite{Hoffmann1985}. Calculations in 1144 compounds showed consistent results for As-As bonding in the hcT transitions \cite{Borisov2018}. However, first principles linear muffin-tin orbital band calculations on SrRh$_2$P$_2$ indicate that P-P bonding is more likely due to the filling of a P-P bonding band which also has $M$-P antibonding character, and the simultaneous filling of an $M$-$M$ antibonding band \cite{Johrendt1997}. 

In contrast to Ca(Fe$_{1-x}$Co$_x$)$_2$As$_2$ and CaKFe$_4$As$_4$ where bulk superconductivity is entirely suppressed when the system undergoes the transition to cT or hcT, Sr(Ni$_{1-x}$Rh$_x$)$_2$P$_2$ shows superconductivity with comparable $T_c$ on both sides of the structural transition, more resembling other systems like BaBi$_3$ \cite{Xiang2018} and PbTaSe$_2$ \cite{Kaluarachchi2017b}. In iron-based superconductors, magnetic fluctuations have been argued to be involved in the mechanism for superconductivity, and their drastic suppression as the structure undergoes the collapsed (or half-collapsed) tetragonal transition is consistent with the absence of bulk superconductivity in the cT (or hcT) phase. The case of Sr(Ni$_{1-x}$Rh$_x$)$_2$P$_2$ seems to be different in that the change in the magnetic properties between the tcO and ucT phases is very subtle, and the changes in the superconducting properties may be governed by other parameters typically relevant in BCS superconductors.

\begin{figure}[H]
 \centering
\includegraphics[width=\linewidth]{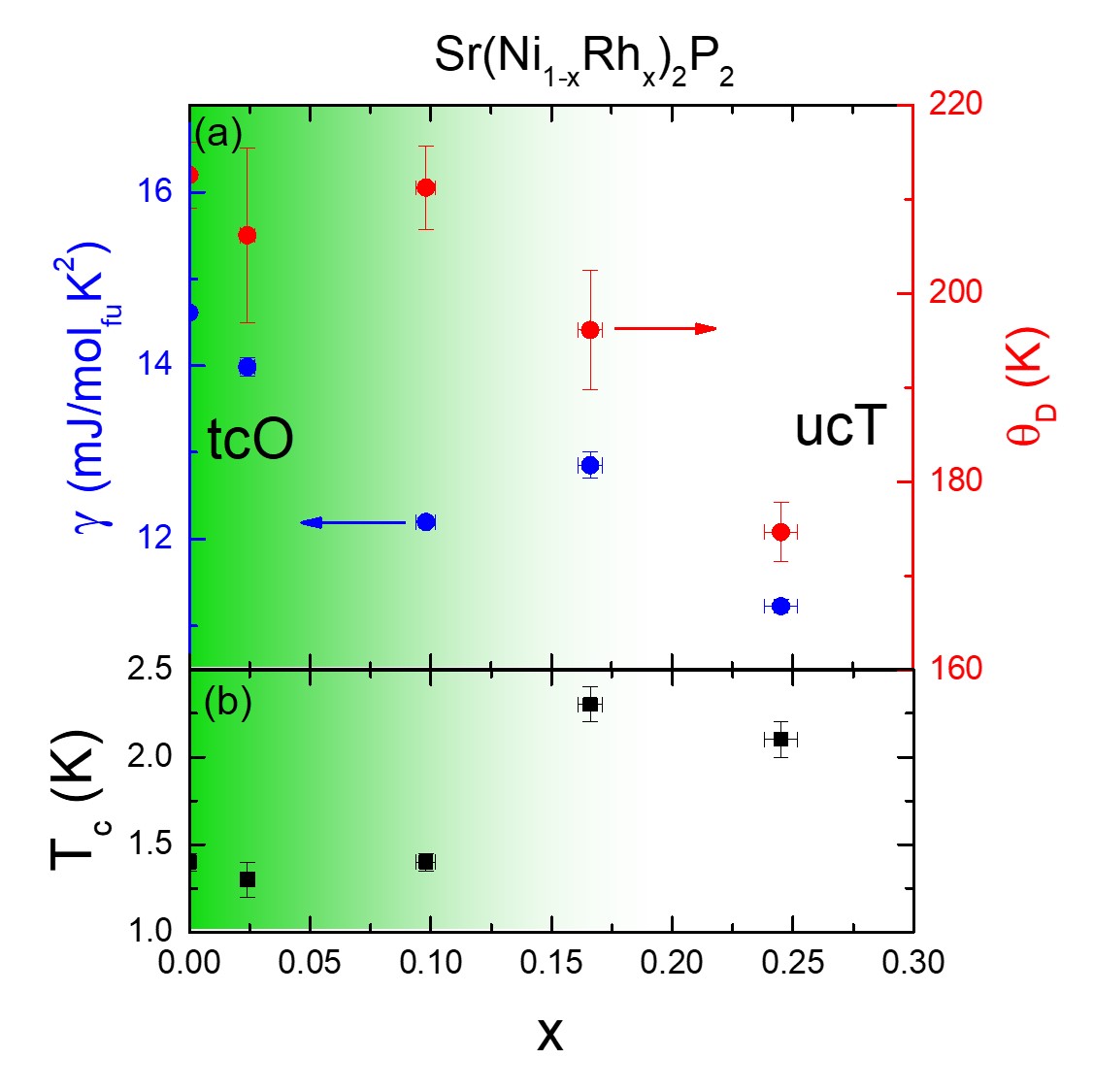}
 \caption{\footnotesize{(a) Sommerfeld coefficient (blue symbols, left axis) and Debye temperature (red symbols, right axis) as estimated from $C_p$ measurements for Sr(Ni$_{1-x}$Rh$_x$)$_2$P$_2$ as a function of the Rh fraction $x$. (b) Superconducting transition temperature of Sr(Ni$_{1-x}$Rh$_x$)$_2$P$_2$ as a function of the Rh concentration $x$, extracted from $R(T)$ measurements for selected concentrations. The region shaded with green corresponds to the compositions with a tcO groundstate.}}
 \label{fig:cp_param}
\end{figure}

In the simplest model of a BCS superconductor, the value of $T_c$ can be expressed in terms of the Debye temperature, $\theta_D$, and density of states at the Fermi level, $n(\varepsilon_F)$, and the phonon-mediated interaction strength, $V$, by
\begin{equation}
    T_c\approx 1.13\theta_D\ e^{-\frac{1}{n(\varepsilon_F)V}}.
    \label{eq:Tc}
\end{equation}
The value of $\theta_D$ can be estimated from the specific heat measurements presented in Fig. \ref{fig:cp}, by means of the expression $\beta=234 k_B N_A \theta_D^{-3}$,
where $\beta$ corresponds to the slope of the linear fit done in Fig. \ref{fig:cp}. The density of states is proportional to the Sommerfeld coefficient, $\gamma$, obtained from the intercept of that linear fit. The dependence of $\theta_D$ and $\gamma$ with the Rh fraction are plotted in Fig. \ref{fig:cp_param}(a). 

Given the overall decrease in both 
$\theta_D$ and $\gamma$, it seems most likely that the tcO$\rightarrow$ucT transition enhances $T_c$ through an enhancement of the electron-phonon coupling strength $V$, which is the only independent variable left in eq. \ref{eq:Tc}. This could happen as a consequence of the breaking of the P-P bonds in the ucT phase. Indeed, the opposite has been observed when the P-P bonds are formed: a decrease in $T_c$ of SrNi$_2$P$_2$ under pressure has been associated to a transition from the tcO to the cT state \cite{Ronning2009}. Further studies, such as inelastic scattering experiments focused on measuring phonon spectra or detailed DFT calculations could provide further insight into the changes electron-phonon coupling strength, and ultimately the relationship to superconductivity in these materials.

\begin{table}
\caption{\label{tab:phase_diagram} Superconducting and structural transition temperatures of Sr(Ni$_{1-x}$Rh$_x$)$_2$P$_2$ for different values of $x$, obtained from temperature-dependent resistance, magnetization and x-ray diffraction measurements. The lack of $T_c$ values for $x\leq0.122$ (with the exception of $x=0$, 0.024 and 0.098) does not indicate a lack of superconductivity, but rather that measurements were not performed below 1.8 K where superconductivity is expected.}
\begin{ruledtabular}
\begin{tabular}{c|ccc|ccc}
 &\multicolumn{2}{c}{$T_c$ (K)}&\multicolumn{3}{c}{$T_S$ (K)}\\
 $x$&from  & from & from 
&from &from &from\\
 &$R(T)$ &$M(T)$ &$C_p(T)$ &$R(T)$
&$M(T)$ &XRD\\
  \hline
 0& 1.4(1) & & & 322(5) & 323(3) & 320(5) \\
  
0.005(1)& & & &314(4) &312(3) & \\
   
0.012(1)& & & &300(5) & 303(3) & \\
      
0.024(2)& 1.3(1) &  & & 279(5) & 276(5) & 276(4) \\
      
0.036(2)& & & & 252(5) & 261(10) & \\
      
0.055(5)& & & & 218(10) & 214(10) & \\
      
0.098(4)& 1.4(1) & & & 157(5) & 153(5) & 150(10)\\
            
0.122(6)& 1.6(4) &  & & 81(15) &  & \\

0.166(5)& 2.3(1) & 2.3(1) & 2.4(1) & & & \\
      
0.195(6)& 2.26(5) & 2.3(1) & & & & \\

0.245(7)& 2.1(1) & 2.1(1) & 2.1(1) & & & \\
    
\end{tabular}
\end{ruledtabular}

\end{table}

\subsection{Effects of Rh substitution on the mechanical properties}

\begin{figure}[H]
 \centering
 \includegraphics[width=\linewidth]{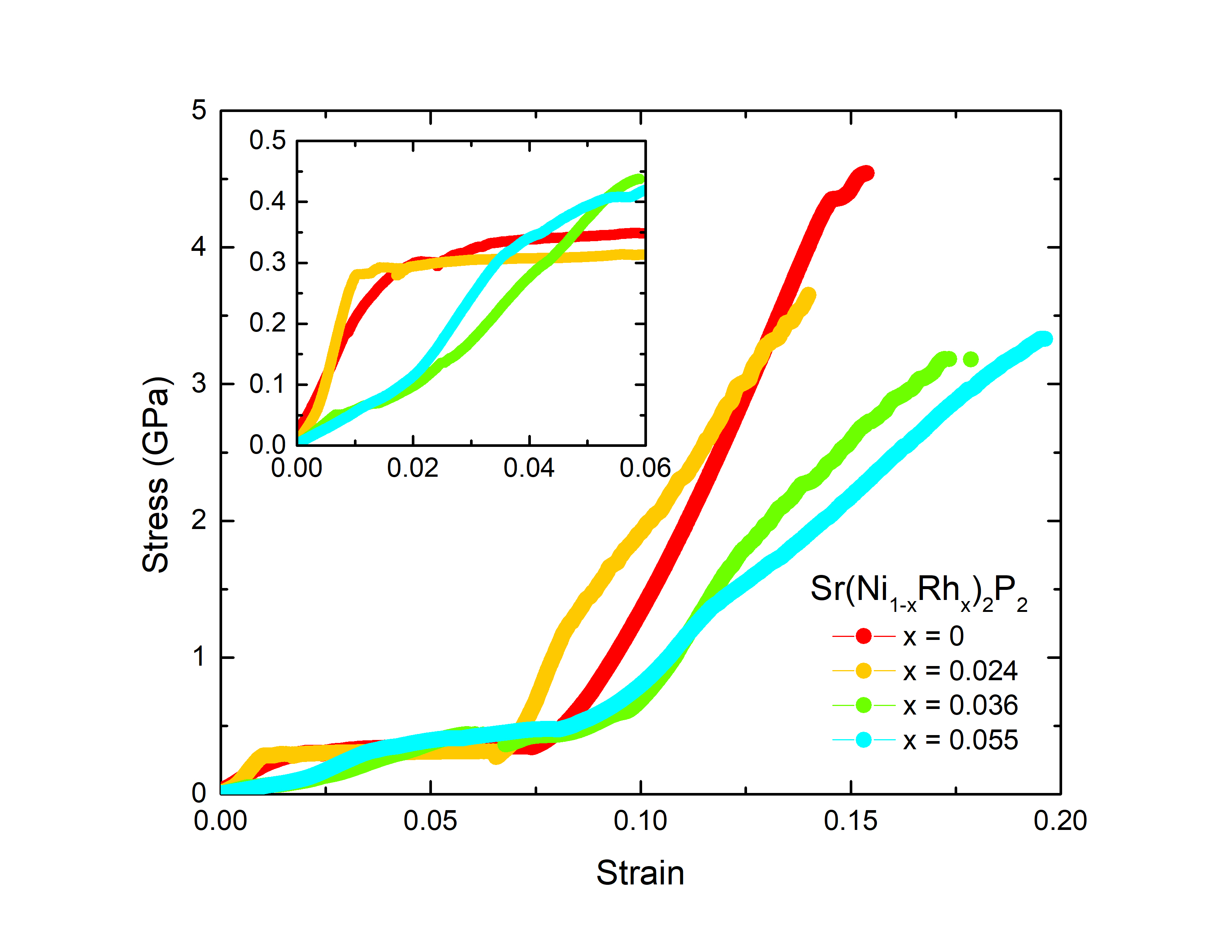}
 \caption{\footnotesize{Main panel: Room temperature uniaxial compressive stress applied along the $c$-axis vs strain along the $c$-axis, for samples of Sr(Ni$_{1-x}$Rh$_x$)$_2$P$_2$ with different values of $x$ as indicated by the different colors. Inset: Enlarged view of the lower strain and stress region.}}
 \label{fig:uniaxialstress}
\end{figure}

In order to provide some insight on the effect of stress on the structural properties of this system, room temperature stress-strain curves were obtained for Sr(Ni$_{1-x}$Rh$_x$)$_2$P$_2$ with $x=0$, $x=0.024$, $x=0.036$ and $x=0.055$, shown in Fig. \ref{fig:uniaxialstress}. Compressive uniaxial stress was applied along the $c$ axis of the crystals in a manner similar to references \citenum{Xiao2021,Song2019,Sypek2017}. Fig. \ref{fig:uniaxialstress} shows that Rh substitution affects the collapse transitions induced by uniaxial stress. The plateaus displayed by these curves indicate the presence of a stress-induced phase transition, where a significant change in strain is observed with a relatively smaller increase of the applied stress.   

The clearest change upon Rh substitution can be observed in the inset of Fig. \ref{fig:uniaxialstress}. For those compositions where the ambient pressure tcO phase transition is suppressed far enough below room temperature (i.e. 252 K for $x=0.036$ in green and 218 K for $x=0.055$ in cyan), two plateau-like features can be easily identified: one at stresses below 0.15 GPa that can be associated with the stress-induced ucT$\leftrightarrow$tcO transition, as well as a plateau at stresses above 0.3 GPa associated with the tcO$\leftrightarrow$cT transition. For those compositions where the ambient pressure tcO transition is above or close to room temperature (i.e. 322 K for $x=0$ in red or 279 K for $x=0.024$ in orange) there is only one clearer plateau around 3 GPa that is associated with the transition from tcO to cT and a small plateau (just right before the tcO$\leftrightarrow$cT transition) has been associated with small amounts of ucT phase \cite{Xiao2021} in the samples milled with FIB.

In general qualitative terms, Fig. \ref{fig:uniaxialstress} shows that increasing Rh fraction shifts the tcO$\leftrightarrow$cT to higher critical uniaxial stresses and strains and makes the ucT$\leftrightarrow$tcO transition appear. Moreover, the measurements for Fig. \ref{fig:uniaxialstress} were carried out by increasing the applied uniaxial stress from zero to the maximum that the pillars could take before a plastic slip or a fracture, which is a reasonable estimate of the maximum recoverable strain of the material. As can be appreciated from the figure, there is a modest increase in that maximum strain from 15\% in the sample with $x=0$ (in red) to almost 20\% in the sample with $x=0.055$ (in cyan). 

\section{Conclusion}
\label{sec:Conclusion}

The structural ucT$\leftrightarrow$tcO transition temperature, $T_S$, decreases upon increasing the Rh fraction in Sr(Ni$_{1-x}$Rh$_x$)$_2$P$_2$ until full suppression for $x\geq 0.166$. For the latter, the structure remains in the ucT state down to 1.8 K. The transition temperatures extracted from temperature-dependent resistance and magnetization measurements were consistent with those observed by x-ray diffraction.

Strain-stress characterization of Sr(Ni$_{1-x}$Rh$_x$)$_2$P$_2$ suggest that stress can induce ucT$\leftrightarrow$tcO as well as tcO$\leftrightarrow$cT transitions. The fact that the stress at which these transitions occur increases with increasing the Rh fraction is consistent with the stabilization of the ucT phase with Rh substitution, and hence with the decrease of the structural transition temperature as seen by means of temperature dependent resistance, magnetization and single crystal x-ray diffraction measurements.

Rh substitution in Sr(Ni$_{1-x}$Rh$_x$)$_2$P$_2$ also affects the superconducting properties. The superconducting transition temperature, $T_c$, remains constant and below 1.4 K for $x < 0.122$, and is enhanced to 2.3 K after the tcO phase is fully suppressed for $x=0.166$. For higher $x$ the superconducting critical temperature is suppressed. The specific heat results suggest that the sudden change of $T_c$ that occurs between $x=0.098$ and $x=0.166$ is dominated by a change in the strength of the electron-phonon interaction as the low temperature structure changes from tcO to ucT. On the other hand, the observed superconductivity in the ucT phase is consistent with previous predictions of superconductivity induced by Ni substitution in SrRh$_2$P$_2$ \cite{Johrendt1997}. In this picture, Ni susbstitution would be serving as a way to increase the Fermi energy of the non-superconducting SrRh$_2$P$_2$ towards a peak in the density of states attributed to $M$-$M$ antibonding bands, similar to that present in LuNi$_2$B$_2$C.

\begin{acknowledgements}
The authors would like to thank Guilherme Gorgen-Lesseux for growing several single crystals used in this work. We appreciate the opportunity of
beta-testing of the ADR module, when some of the data were taken. Work done at Ames National Laboratory was supported by the U.S. Department of Energy, Office of Basic Energy Science, Division of Materials Sciences and Engineering. Ames National Laboratory is operated for the U.S. Department of Energy by Iowa State University under Contract No. DE-AC02-07CH11358.

\end{acknowledgements}

\section*{Appendix A}
\begin{table}[H]
\caption{\label{tab:scxrd_SrNi2P2}.}
\begin{ruledtabular}
\begin{tabular}{ll}
\textrm{Chemical formula}&
\textrm{SrNi$_2$P$_2$}\\
\colrule
Formula Weight & 266.95\\
Temperature & 294(9) K\\
Crystal system & orthorhombic\\
Space group & $Immm$\\
 & $a=3.9584(2)\ \text{\AA}$, $\alpha=90^{\circ}$ \\
Unit cell dimensions & $b=11.8763(6)\ \text{\AA}$, $\beta=90^{\circ}$\\
 & $c=10.4778(5)\ \text{\AA}$, $\gamma=90^{\circ}$ \\
Volume & 492.57(4) $\text{\AA}^3$ \\
$Z$ & 6 \\
Calculated density & 5.400 g/cm$^3$\\
Absorption coefficient & 14.930 mm$^{-1}$\\
$F(000)$ & 744.0\\
Crystal size & $0.096\times 0.052\times 0.027$ mm$^3$\\
Radiation & Ag K$\alpha$ ($\lambda=0.56087\ \text{\AA}$)\\
$2\Theta$ range for collection & 5.414$^{\circ}$ to 63.874$^{\circ}$\\
Min/Max index [$h$, $k$, $l$] & [$-$6/7, $-$20/22, $-$18/17]\\
Reflections collected & 1505\\
Independent reflections & 1505[$R_{int}=0.0320$]\\
Data/restraint/parameters & 1505/0/29\\
Goodness-of-fit on $F^2$ & 0.980\\
Final $R$ indexes [$I\geq 2\sigma (I)$] & $R_1=0.0239$, $wR_2=0.0599$\\
Final $R$ indexes [all data] & $R_1=0.0336$, $wR_2=0.0623$\\
Largest diff. peak/hole & 1.60 $e^- \text{\AA}^{-3}$/$-$1.37 $e^- \text{\AA}^{-3}$\\
\end{tabular}
\end{ruledtabular}
\end{table}

\begin{table}[H]
\caption{\label{tab:scxrd_doped1}.}
\begin{ruledtabular}
\begin{tabular}{ll}
\textrm{Chemical formula}&
\textrm{Sr(Ni$_{0.976}$Rh$_{0.024}$)$_2$P$_2$}\\
\colrule
Formula Weight & 269.19\\
Temperature & 297.5(5) K \\
Crystal system & tetragonal\\
Space group & $I4/mmm$\\
 & $a=3.9523(1)\ \text{\AA}$, $\alpha=90^{\circ}$\\
Unit cell dimensions & $b=3.9523(1)\ \text{\AA}$, $\beta=90^{\circ}$\\
 & $c=10.7508(5)\ \text{\AA}$, $\gamma=90^{\circ}$\\
Volume & 167.93(1) $\text{\AA}^3$\\
$Z$ & 2\\
Calculated density & 5.324 g/cm$^3$\\
Absorption coefficient & 14.597 mm$^{-1}$\\
$F(000)$ & 250.0\\
Crystal size & $0.092\times0.082\times0.024$ mm$^3$ \\
Radiation & Ag K$\alpha$ ($\lambda=0.56087\ \text{\AA}$)\\
$2\Theta$ range for collection & 5.98$^{\circ}$ to 62.958$^{\circ}$\\
Min/Max index [$h$, $k$, $l$] & [$-$6/7, $-$7/6,  $-$16/19] \\
Reflections collected & 1951\\
Independent reflections & 187[$R_{int}=0.0377$]\\
Data/restraint/parameters & 187/0/8 \\
Goodness-of-fit on $F^2$ & 1.236\\
Final $R$ indexes [$I\geq 2\sigma (I)$] & $R_1=0.0157$, $wR_2=0.0420$\\
Final $R$ indexes [all data] & $R_1=0.0160$, $wR_2=0.0421$\\
Largest diff. peak/hole & 0.73 $e^- \text{\AA}^{-3}$/$-$9.96 $e^- \text{\AA}^{-3}$ \\

\end{tabular}
\end{ruledtabular}
\end{table}

\begin{table}[H]
\caption{\label{tab:scxrd_doped2}.}
\begin{ruledtabular}
\begin{tabular}{ll}
\textrm{Chemical formula}&
\textrm{Sr(Ni$_{0.902}$Rh$_{0.098}$)$_2$P$_2$}\\
\colrule
Formula Weight & 275.82\\
Temperature & 300.0(1) K\\
Crystal system & tetragonal\\
Space group & $I4/mmm$\\
 & $a=3.9532(2)\ \text{\AA}$, $\alpha=90^{\circ}$\\
Unit cell dimensions & $b=3.9532(2)\ \text{\AA}$, $\beta=90^{\circ}$\\
 & $c=10.7861(9)\ \text{\AA}$, $\gamma=90^{\circ}$\\
Volume & 168.56(2) $\text{\AA}^3$\\
$Z$ & 2\\
Calculated density & 5.434 g/cm$^3$\\
Absorption coefficient & 14.543 mm$^{-1}$\\
$F(000)$ & 255.0\\
Crystal size & $0.114\times0.105\times0.032$ mm$^3$\\
Radiation & Ag K$\alpha$ ($\lambda=0.56087\ \text{\AA}$)\\
$2\Theta$ range for collection & 5.962$^{\circ}$ to 62.974$^{\circ}$\\
Min/Max index [$h$, $k$, $l$] & [$-$7/5, $-$6/6, $-$17/15]\\
Reflections collected & 1152\\
Independent reflections & 177[$R_{int}=0.0223$]\\
Data/restraint/parameters & 177/0/8\\
Goodness-of-fit on $F^2$ & 1.167\\
Final $R$ indexes [$I\geq 2\sigma (I)$] & $R_1=0.0197$, $wR_2=0.0454$\\
Final $R$ indexes [all data] & $R_1=0.0205$, $wR_2=0.0458$\\
Largest diff. peak/hole & 0.89 $e^- \text{\AA}^{-3}$/$-$1.02 $e^- \text{\AA}^{-3}$\\

\end{tabular}
\end{ruledtabular}
\end{table}

\begin{table}[H]
\caption{\label{tab:scxrd_SrRh2P2}.}
\begin{ruledtabular}
\begin{tabular}{ll}
\textrm{Chemical formula}&
\textrm{SrRh$_2$P$_2$}\\
\colrule
Formula Weight & 355.38\\
Temperature & 297.9(1) K\\
Crystal system & tetragonal\\
Space group & $I4/mmm$\\
 & $a=3.9371(2)\ \text{\AA}$, $\alpha=90^{\circ}$\\
Unit cell dimensions & $b=3.9371(2)\ \text{\AA}$, $\beta=90^{\circ}$\\
 & $c=11.715(1)\ \text{\AA}$, $\gamma=90^{\circ}$\\
Volume & 181.59(3) $\text{\AA}^3$\\
$Z$ & 2\\
Calculated density & 6.499 g/cm$^3$\\
Absorption coefficient & 12.862 mm$^{-1}$\\
$F(000)$ & 316.0\\
Crystal size & $0.08\times 0.07\times 0.05$ mm$^3$\\
Radiation & Ag K$\alpha$ ($\lambda=0.56087\ \text{\AA}$)\\
$2\Theta$ range for collection & 5.488$^{\circ}$ to 63.168$^{\circ}$\\
Min/Max index [$h$, $k$, $l$] & [$-$7/6, $-$5/7, $-$19/12]\\
Reflections collected & 628\\
Independent reflections & 187[$R_{int}=0.0173$]\\
Data/restraint/parameters & 187/0/8\\
Goodness-of-fit on $F^2$ & 1.101\\
Final $R$ indexes [$I\geq 2\sigma (I)$] & $R_1=0.0243$, $wR_2=0.0559$\\
Final $R$ indexes [all data] & $R_1=0.0243$, $wR_2=0.0559$\\
Largest diff. peak/hole & 1.60 $e^- \text{\AA}^{-3}$/$-$1.38 $e^- \text{\AA}^{-3}$\\

\end{tabular}
\end{ruledtabular}
\end{table}

\section*{Appendix B}

In this appendix we compare the results of powder and single crystal x-ray diffraction, in order to provide an explanation as to why the coexistence of the ucT and the tcO phases is seen in the former, even far beyond the hysteresis region observed with temperature dependent resistance measurements, with the exception of $x=0.012(1)$ for which the transition occurs close enough to room temperature. For this particular composition, coexistence at room temperature is expected, since the hysteresis region extends between 295 K and 305 K. In addition, small inhomogeneities of Rh content could be enough to stabilize both ucT and tcO phases on the same sample. However, this does not apply to any of the other compositions in which coexistence was detected in the powdered samples. In this appendix, we provide an alternative explanation to this.

\begin{figure}[H]
\centering
\includegraphics[width=\linewidth]{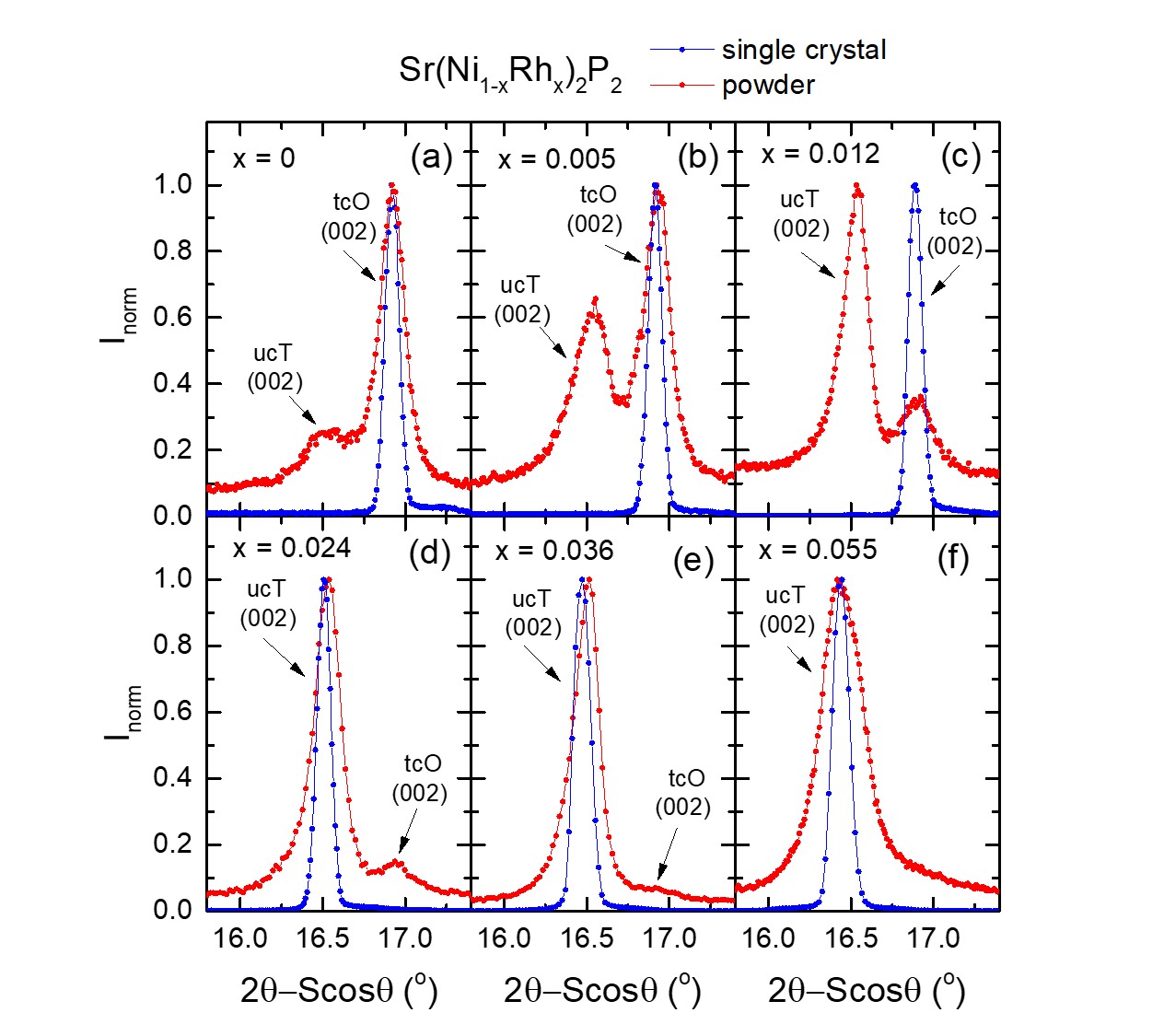}
\caption{\footnotesize{Profile of the (002) peaks measured on single-crystalline plates (blue) and powder (red) using the same reflection mode Bragg-Brentano geometry. The results are plotted as a function of $2\theta - S \cos (\theta)$ in order to account for the different vertical displacements of the irradiated sample surfaces, according to reference \citenum{Jesche2016}}.}
\label{fig:plate_powder}
\end{figure}

 The tcO and ucT phases are distinguishable from each other due to the fact that they belong to different space groups: $Immm$ and $I4/mmm$, respectively. Particularly, tcO phase exhibits \textit{satellite} peaks that are forbidden for the ucT due to the latter's higher symmetry. In the powder x-ray diffraction measurements, many of the peaks of the tcO phase overlap with those of the ucT phase due to their similar $a$-lattice parameter. However, for some compositions it is possible to resolve two distinct peaks corresponding to the indices $(00l)$, as is the case for the $(002)$ peaks shown in detail in Fig. \ref{fig:plate_powder}.

In contrast with the x-ray diffraction measurements done on ground powders, the measurements done on single crystals do not show any coexistence of the ucT and tcO phases at all. This can be clearly demonstrated by the measurements done on single-crystalline plates with reflection mode Bragg-Brentano geometry of the Rigaku MiniFlex II powder diffractometer. Fig. \ref{fig:plate_powder} shows the $(0 0 2)$ peaks measured on single crystals (in blue) compared to the results measured on the powder (in red), for $x\leq 0.055$. For a better comparison between the powder and single crystal x-ray patterns, the results are plotted as a function of $2\theta - S \cos{\theta}$, where $S$ is a geometrical factor that accounts for the different vertical displacement of the irradiated sample's surface, which can be calculated following the procedure described in reference \citenum{Jesche2016}. It should be noted that the samples used for these measurement can have an arbitrary large area, in our case being larger than 2 mm by 2 mm for all of the samples. The fact that there are only one set of $(002)$ peaks for the single crystals indicates a single phase across the whole area of these samples, and suggests that the observed phase coexistence in the powder is induced as a result of the grinding, which extends far beyond the hysteresis region reflected in the temperature dependent resistance measurements.

It is worth noting that the single crystal data shown in blue in Fig. \ref{fig:plate_powder} are consistent with the resistance data shown in Fig. \ref{fig:rt} and the phase diagram data shown in Fig. \ref{fig:phase_diagram}. At room temperature, for $x=0$, 0.005 and 0.012, the single crystal data show the (002) peak associated with the tcO phase and for higher $x$-values it shows the (002) peak associated with the ucT phase. It is also worth noting that as the tcO transition temperature drops further below room temperature, the size of the tcO peak observed in the powder data diminishes.

The single crystal x-ray diffraction results obtained with the Rigaku XtaLab Synergy-S diffractometer in transmission mode were also consistent with a single phase in all the crystals measured. 

\begin{table}
\caption{\label{tab:latticeparam} $a$ and $c$-lattice parameters of Sr(Ni$_{1-x}$Rh$_x$)$_2$P$_2$ for different values of $x$, and measured on powder and single crystals in reflection mode, as well as on single crystals measured on transmission mode.}
\begin{ruledtabular}
\begin{tabular}{c|cc|ccc}
 &\multicolumn{2}{c}{$a$ (\text{\AA})}&\multicolumn{3}{c}{$c$ (\text{\AA})}\\
 $x$&powder&single&powder
&single&single\\
 & &crystal\footnote{Transmission mode}& &crystal$^a$ &crystal\footnote{Reflection mode (Bragg-Brentano)}\\ \hline
 0&3.949 &  &10.734 &  & \\
  &3.959 &3.9584 &10.481 & 10.4778&10.481 \\
  \hline
  
0.005&3.952 & &10.731 & & \\
  &3.961 & 3.9613 &10.49 & 10.4866&10.486 \\
\hline
   
0.012&3.950 & &10.73 & & \\
  &3.964 & 3.9616 &10.49 & 10.4976 &10.493 \\
\hline
      
0.024&3.952 &3.9523 &10.75 &10.7511 & 10.740\\
  &3.974 & &10.48 & & \\
\hline
      
0.036&3.9469 & 3.9468 &10.746 &10.751 & 10.756\\
  &3.982 & &10.462 & & \\
\hline
      
0.055&3.9474 & 3.9518 &10.767 & 10.787 &10.779 \\
  &3.9841 & &10.458 & & \\
\hline
      
0.098&3.9543 &3.9532 &10.7780 &10.7861 & 10.796 \\
  & & & & & \\
 \hline
      
0.122&  & 3.9547 &  & 10.820 & 10.824 \\
  & & & & & \\
 \hline
      
0.166&3.9544 & 3.9557 &10.871 & 10.870& 10.873\\
  & & & & & \\
\hline
      
0.195&3.9547 & 3.9567&10.895 & 10.8919 & 10.89\\
  & & & & & \\
\hline

0.245&3.9569 & 3.9553 &10.9323 & 10.926 &10.947 \\
  & & & & & \\
\hline

1&  &3.9371 & &11.715 & \\
& & & & & \\
    
\end{tabular}
\end{ruledtabular}

\end{table}

\begin{figure}[tbh]
\centering
\includegraphics[width=0.85\linewidth]{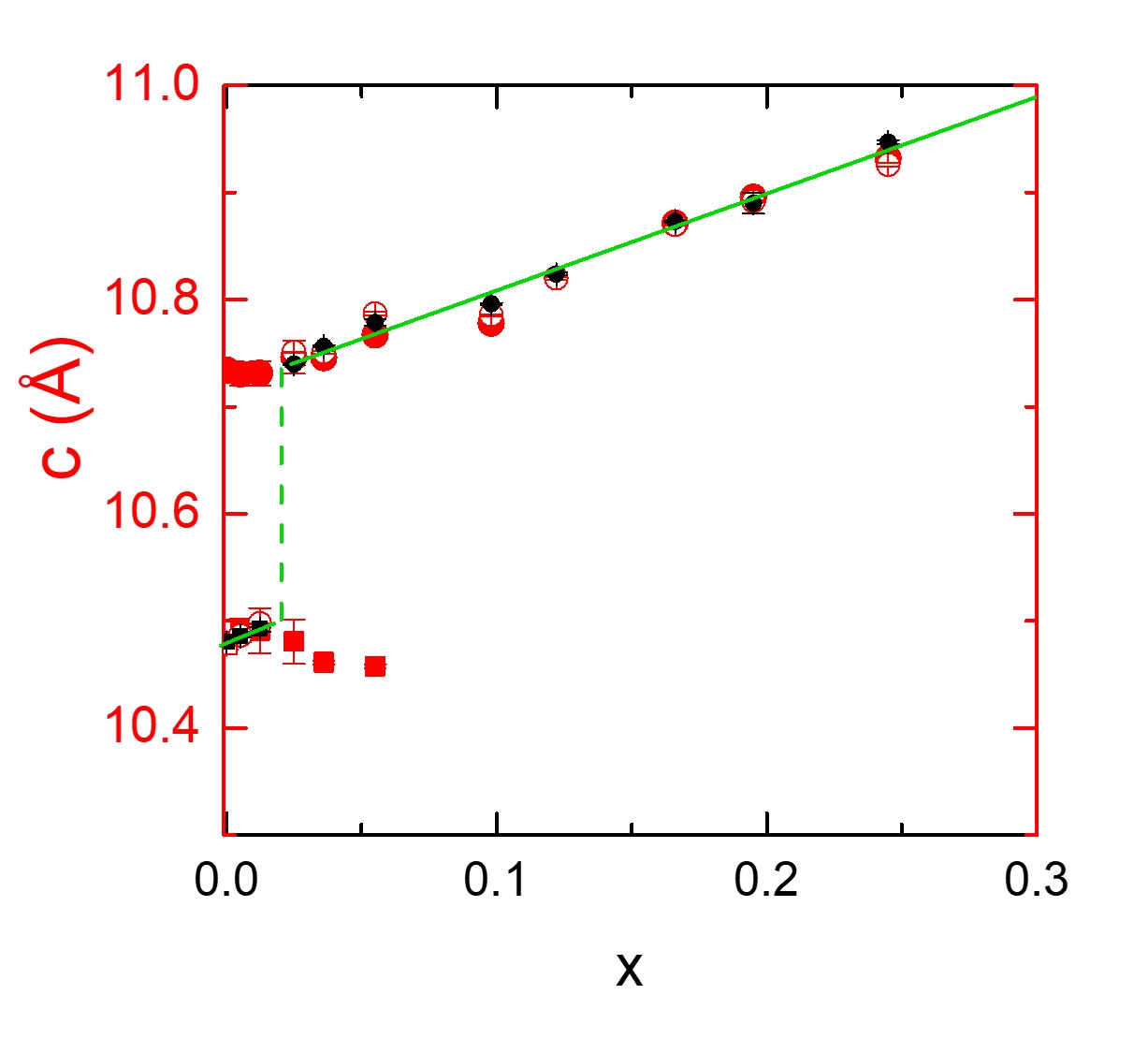}
\caption{\footnotesize{$c$-lattice parameters of Sr(Ni$_{1-x}$Rh$_x$)$_2$P$_2$ as a function of $x$, and measured on powder (solid red symbols) and single crystals in reflection mode (solid black symbols), as well as on single crystals measured in transmission mode (open red symbols); those corresponding to the ucT phase are represented with circles, and those corresponding to the tcO phase are represented in squares. A green line is used as a guide to the eye to show the intrinsic trend in the $c$-lattice parameters.}}
\label{fig:latticeparam}
\end{figure}

The room temperature lattice parameters measured through each technique (powder with the Bragg-Brentano geometry, single crystal in reflection mode, and single crystal in transmission mode) were all consistent with each other. Surprisingly, this includes the lattice parameters observed in one of the phases in the powder (even though one could have expected grinding to change the lattice parameters as well), and excludes the lattice parameters of the other phase induced by the grinding. This is shown in Table \ref{tab:latticeparam} as well as in Fig. \ref{fig:latticeparam}, where the open symbols correspond to the measurements done on single crystals in transmission mode, and the solid symbols correspond to the measurements done on powder; the circles correspond to the ucT phase, and the squares to the tcO phase. On top of this, the $c$-lattice parameters obtained from measuring the single crystals in reflection mode with the Bragg-Brentano geometry were added to the plot as solid black symbols.

\section*{Appendix C}

\begin{figure}[H]
 \centering
 \includegraphics[width=\linewidth]{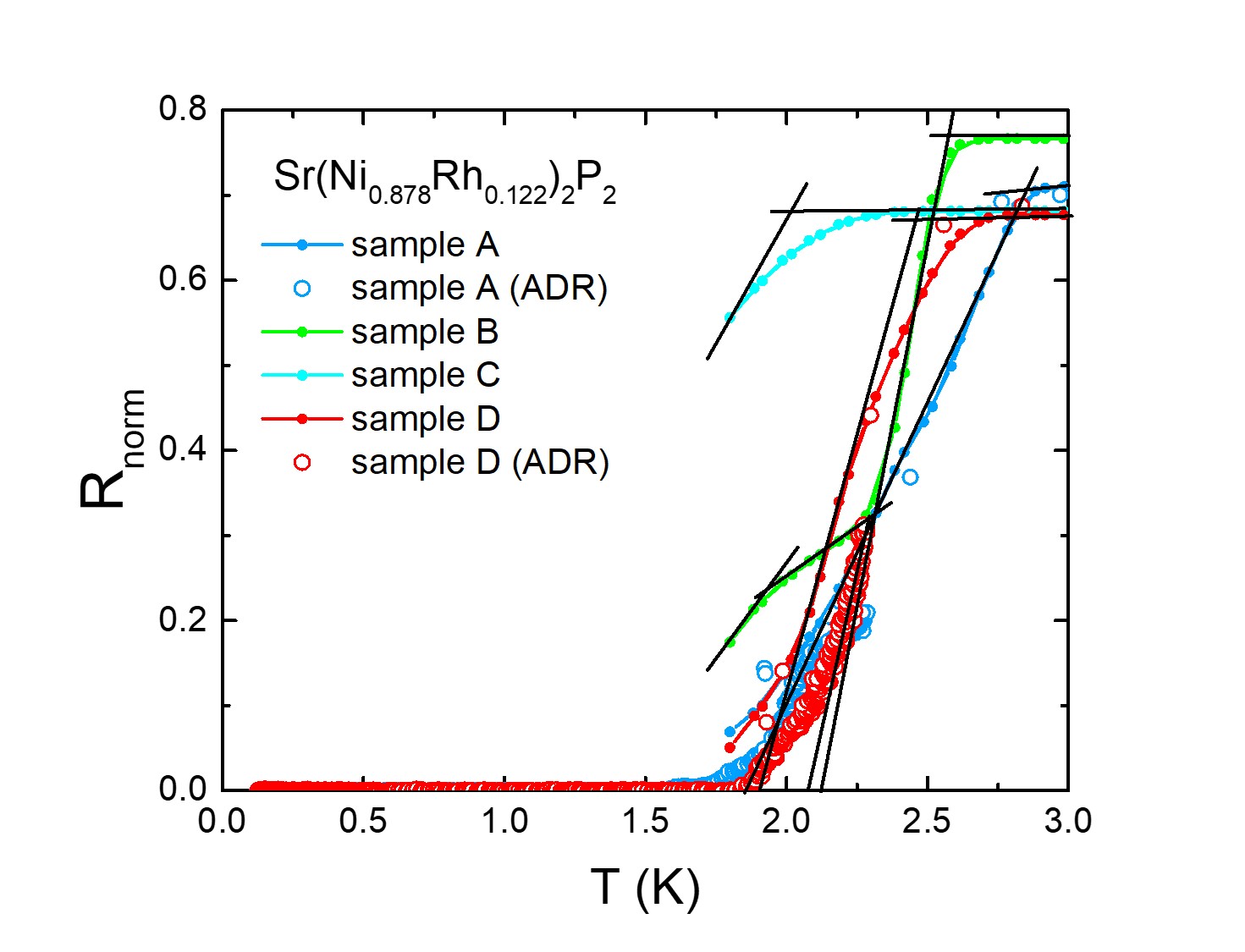}
 \caption{\footnotesize{Enlarged view of the temperature dependent resistance for four different samples of Sr(Ni$_{1-x}$Rh$_x$)$_2$P$_2$ with $x \approx 0.122$. The black lines indicate the lines of maximum slope and the extrapolated high temperature behavior, used to estimate the onset and offset temperatures of the superconducting transition. The open symbols correspond to measurements done with ADR.}}
 \label{fig:RT_ZX312}
\end{figure}

As mentioned in Section C, for $x=0.122$ the samples exhibit a broader superconducting transition with a more complicated shape than other values of $x$, as well as a greater sample-to-sample variation in the onsets and offsets of superconductivity. This can be seen in Fig. \ref{fig:RT_ZX312} for the four different samples with $x\approx 0.122$ in which resistance was measured. The onset and offset temperatures of the superconducting transition were estimated using the same criteria as for the rest of the samples, by extrapolating the line of maximum slope to intersect the high temperature behavior or the low temperature zero resistance, respectively. For the particular case sample C (in cyan) there was not enough low temperature data points, so only the onset was estimated. For the case of sample B, since it displayed a double step behavior, the onset and the offset was estimated for the higher temperature step, and only the onset was estimated for the lower temperature step. It should be noted that Fig. \ref{fig:phase_diagram} is missing two offset temperatures that could not be determined due to lack of lower temperature points for those two samples.

\begin{figure}[H]
 \centering
 \includegraphics[width=\linewidth]{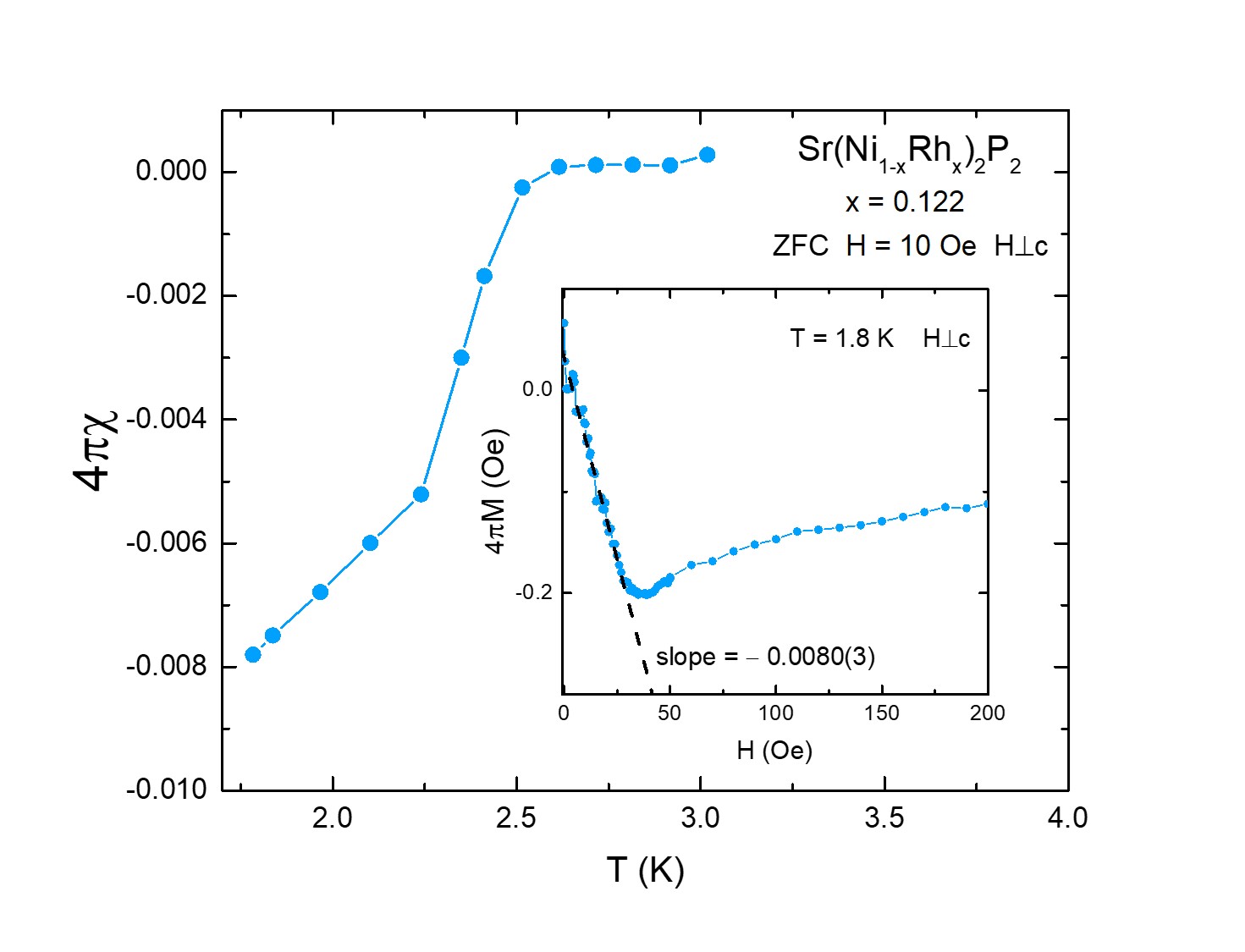}
 \caption{\footnotesize{Main panel: magnetic susceptibility as a function of temperature for Sr(Ni$_{1-x}$Rh$_x$)$_2$P$_2$ with $x = 0.122$. Inset: Magnetization as a function of magnetic field for the former compositions. The small value of the slope indicates this composition indicates that the observed diamagnetic signal is likely due to an impurity that represents less than 1\% of the volume of the sample.}}
 \label{fig:Mag_sc_zx312}
\end{figure}

The magnetic susceptibility, $\chi$, is presented in Fig. \ref{fig:Mag_sc_zx312} for another $x\sim 0.122$ sample, showing that $4\pi\chi\approx -0.0080(3)$ at $T=1.8\ \text{K}$, which is consistent with an expulsion of the magnetic field in less than 1\% of the sample at that temperature. This suggests that the transitions observed for $R(T)$ measurements in Fig. \ref{fig:RT_ZX312} are likely due to a small fraction of the sample in a metastable ucT phase, such that the $T_c$ is enhanced for that fraction of the sample only. This is consistent with the $x=0.122$ sample being very close to the point where the tcO transition is suppressed to 0 K, as shown in the phase diagram of Fig. \ref{fig:phase_diagram}.

\newpage 

\label{sec:Appendix}

\nocite{apsrev41Control}
\bibliographystyle{apsrev4-1}
\bibliography{CollapsedTet.bib}

\end{document}